\documentclass[twocolumn]{aastex701}

\usepackage{xspace}
\usepackage{booktabs}  %\addlinespace[]
\usepackage{amsmath, amstext}   %\begin{aligned}
\hypersetup{colorlinks, linkcolor={blue}, citecolor={blue}, urlcolor={blue}}

\newcommand\rxte{\textit{RXTE}\xspace}
\newcommand\ixpe{\textit{IXPE}\xspace}

\newcommand\nicer{\textit{NICER}\xspace}
\newcommand\chandra{\textit{Chandra}\xspace}
\newcommand\swift{\textit{Swift}\xspace}
\newcommand\xmm{\textit{XMM-Newton}\xspace}

\newcommand\targeta{1E~2259+586\xspace}
\newcommand\targetb{4U~0142+61\xspace}
\newcommand\targetc{1RXS~J170849.0-400910\xspace}
\newcommand\targetcc{1RXS~J1708\xspace}
\newcommand\targetd{1E~1841-045\xspace}
\newcommand\target{Magnetars\xspace}

\begin{document}

\title{Long-term timing evolution of four Anomalous X-Ray Pulsars}

\author[0009-0009-8477-8744]{Han-Long Peng}
\affiliation{School of Physics and Technology, Nanjing Normal University, Nanjing, 210023, Jiangsu, China}
\email{penghl@ihep.ac.cn}

\author[0000-0001-7595-1458]{Shan-Shan Weng}
\affiliation{School of Physics and Technology, Nanjing Normal University, Nanjing, 210023, Jiangsu, China}
\email[show]{wengss@njnu.edu.cn}

\author[0000-0002-2749-6638]{Ming-Yu Ge}
\affiliation{Key Laboratory of Particle Astrophysics, Institute of High Energy Physics, Chinese Academy of Sciences, Beijing 100049, China}
\email[show]{gemy@ihep.ac.cn}

\author[0000-0003-1480-2349]{Shi-Qi Zhou}
\affiliation{School of Physics and Astronomy, China West Normal University, Nanchong 637002, People's Republic of China}
\email{Pulsar.Sq.Zhou@gmail.com}

\author{Erbil G\"{u}gercino\u{g}lu}
\affiliation{School of Arts and Sciences, Qingdao Binhai University, Huangdao District, 266555, Qingdao, China}
\email{egugercinoglu@gmail.com}

\author[0000-0002-1662-7735]{Wen-Tao Ye}
\affiliation{Key Laboratory of Particle Astrophysics, Institute of High Energy Physics, Chinese Academy of Sciences, Beijing 100049, China}
\email{yewt@ihep.ac.cn}

\author[0000-0003-3127-0110]{You-Li Tuo}
\affiliation{Institut f{\"u}r Astronomie und Astrophysik, Universit{\"a}t T{\"u}bingen, Sand 1, D-72076 T{\"u}bingen, Germany}
\email{youli.tuo@astro.uni-tuebingen.de}

\author{Liang Zhang}
\affiliation{Key Laboratory of Particle Astrophysics, Institute of High Energy Physics, Chinese Academy of Sciences, Beijing 100049, China}
\email{zhangliang@ihep.ac.cn}

\author{Juan Zhang}
\affiliation{Key Laboratory of Particle Astrophysics, Institute of High Energy Physics, Chinese Academy of Sciences, Beijing 100049, China}
\email{zhangjuan@ihep.ac.cn}

\author{Shi-Jie Zheng}
\affiliation{Key Laboratory of Particle Astrophysics, Institute of High Energy Physics, Chinese Academy of Sciences, Beijing 100049, China}
\email{zhengsj@ihep.ac.cn}

\author{Yu-Jia Zheng}
\affiliation{School of Physics and Technology, Nanjing Normal University, Nanjing, 210023, Jiangsu, China}
\email{zhengyj@ihep.ac.cn}

\author[0009-0001-7213-2235]{Xian-Ao Wang}
\affiliation{Department of Astronomy, Yunnan University, Kunming 650500, China}
\email{wangxianao1@stu.ynu.edu.cn}

%%%%%%%%%%%%%%%%%%%%%%%%%%%%%%%%%%%%%%%%%%%%%%%%%%%%%%%%%%%%%%%%%%%%%%%%%%%%%%%%%%%%%%%%%%%%%%%%%%
\begin{abstract}

Anomalous X-ray pulsars (AXPs) and soft gamma-ray repeaters (SGRs) are believed to be manifestations of magnetars. Typically, AXPs exhibit higher X-ray luminosities, whereas SGRs are generally fainter and display significantly high signal-to-noise ratios only during their outburst phases. In this work, we report the long-term timing evolution of four AXPs: \targeta, \targetb, \targetc and \targetd,  which were regularly monitored with \nicer from 2017 to 2024. Over this period, we identify a total of 10 timing events. In addition to one glitch and one anti-glitch in \targeta reported in literature,  we detect another 8 new timing events: 5 glitches, 2 anti-glitches, and 1 unusual state transition event. Notably, both anti-glitches were observed in \targetb, making it the most frequent source of such events, and there is a hint of regular evolution in its pulse profile. In the case of \targetc,  it continues to exhibit pronounced high-frequency timing anomalies and undergoes a state transition event. 
Finally, we study the evolution of the pulse profiles and find that the profiles of \targeta and \targetb both evolve. This is consistent with the earlier finding that pulse profile evolution is a generic feature of magnetars.

\end{abstract}


%%%%%%%%%%%%%%%%%%%%%%%%%%%%%%%%%%%%%%%%%%%%%%%%%%%%%%%%%%%%%%%%%%%%%%%%%%%%%%%%%%%%%%%%%%%%%%%%%%
\section{Introduction}

The diagram of rotation period ($P$) versus spin-down rate ($\dot{P}$) is typically employed to classify pulsar populations. Magnetars occupy the upper-right region of the $P-\dot{P}$ diagram ($\mathit{P} \sim 2-12 \, ~\mathrm{s}$ and $\dot{P} \sim  10^{-13}-10^{-10} \, ~\mathrm{s~s^{-1}}$), representing the young neutron stars with the most intense magnetic fields \citep{Olausen_2014}. 
Historically, magnetars first appeared under the names of soft gamma-ray repeaters (SGRs) and anomalous X-ray pulsars (AXPs) \citep{Mazets_a, Fahlman_1981}.
The detection of pulsations in SGRs and SGR-like bursts in AXPs \citep{Mazets_b, Gavriil2002_na} provided strong evidence for the unified magnetar model proposed by \cite{Thompson_1995}. This model attributes the energy sources of both phenomena to the ultra-strong magnetic fields of neutron stars, rather than to spin-down power \citep{Thompson_1995, Thompson_1996}.
As of now, approximately 30 confirmed and candidate magnetars have been cataloged, with a comprehensive and regularly updated listing available in the McGill Magnetar Catalog\footnote{\url{http://www.physics.mcgill.ca/~pulsar/magnetar/main.html}} \citep{Olausen_2014}.

One defining characteristic of magnetars is their rich and diverse radiative behavior, particularly at X-ray and soft gamma-ray energies. These are broadly divided into short-duration explosive events (bursts/flares, including giant flares) and outbursts. The former typically last from milliseconds to minutes, whereas the latter may decay over weeks to months or even years \citep{Kaspi_2017, Esposito_2021, Francesco_2017}. These episodes are often accompanied by timing anomalies \citep{Dib_2008}, such as glitches. Sudden changes in spin frequency are known as glitches, commonly seen in young, rotation-powered pulsars like the Crab pulsar \citep{Lower_2021, Antonopoulou_2022}. Unlike in ordinary rotation-powered pulsars, where glitches are usually discussed in terms of internal superfluid dynamics under an external electromagnetic torque, in magnetars the timing behavior is often intertwined with magnetic-field evolution and magnetically driven crustal activity. Within the magnetar framework, the strong and evolving magnetic field can build up stresses in the crust and induce fractures or plastic flow, which may make magnetars prone to glitches. Such crustal motions can perturb vortex-pinning conditions and the crust--superfluid coupling, triggering collective vortex unpinning and a rapid transfer of angular momentum to the crust. Consistent with this picture, glitches have been observed in a substantial fraction of magnetars in long-term timing campaigns \citep{Dib_2014}.

Most of the glitch events observed in pulsars are spin-up glitches characterized by a sudden increase in the rotational frequency. In addition to these, only a handful of pulsars have experienced more rare glitch events: slow glitches and delayed spin-ups \citep{zhou_2022a}. Since the launch of \rxte and with the help of high resolution power imaging observations from space-borne satellites such as \chandra, \swift, and \xmm, many of the unique timing and radiative behaviors of magnetars have been uncovered. In fact, magnetars have indeed been found to be a type of neutron star that frequently exhibit glitches. Perhaps more intriguingly, beyond ordinary glitches, an unique phenomenon was identified in magnetars: anti-glitches, also known as spin-down glitches, which occur in contrast to spin-up glitches and are characterized by a sudden decrease in rotational frequency. \citep{Archibald_2013}. Although a few anti-glitch events have since been reported in a binary system (NGC 300 ULX-1) and two young, rotation-powered pulsars \citep{Ray_2019, Zhou_2024, Tuo_2024}, nearly all known anti-glitches still occur in magnetars. The underlying reason why magnetars are particularly susceptible to glitches — and especially to anti-glitches -- remains poorly understood. Furthermore, anti-glitches are far less common than normal glitches, and the relationship between the two types of events, as well as the rarity of anti-glitches, continues to be an open question. Two possible physical explanations for anti-glitches are the radial inward motion of superfluid vortex lines driven by crustquakes due to the evolution of the magnetic field within the crust \citep{Akbal_2015}, or the time-variable coupling of magnetospheric and superfluid torques \citep{Erbil_2017b}.

Unlike rotation-powered pulsars, magnetars exhibit rich X-ray activity.  Numerous studies have sought to investigate the connection between radiative activity and glitches, aiming to uncover a possible causal link. Establishing such a relationship would provide critical insights into the triggering mechanisms for magnetar activity and could help constrain theoretical models and geometric configurations of the timing behavior and the magnetic field structure within the crust and in the magnetosphere. For example, \citet{Dib_2014} performed a statistical analysis of five magnetars using 16 years of \rxte data. They found that while most radiative events, such as bursts, flux enhancements, and pulse profile changes, were accompanied by timing anomalies, only $\sim 30\%$ of timing anomalies were associated with observable radiative changes. The overlapping characteristics between spin-up and spin-down glitches, as well as between radiatively silent and active glitch events, and the similarity of AXP glitch behavior to that of the radio pulsar population, suggest a shared internal origin for all glitches, as has long been proposed for rotation-powered pulsars \citep{Anderson_1975, Pines_1985}. However, alternative views exist: some studies have proposed an external-torque interpretation, in which magnetospheric reconfigurations (e.g., changes in the twist/current system, partial opening/restructuring of field lines, or a variable particle wind) lead to abrupt changes in the spin-down torque and can generate timing signatures that may be interpreted as glitch-like events in phase-connected analyses \citep{Lyutikov_2013, Tong_2014}. Notably, \citet{Archibald_2017} reported a net spin-down glitch in \targetb, which followed a decaying spin-up glitch. Their subsequent analysis indicated that nearly all spin-down glitches in magnetars were accompanied by radiative changes, further reinforcing the complexity of the glitch–activity relationship.

To date, numerous observational properties of magnetars have been discovered, and a number of review articles have summarized this progress. Some of them primarily focus on observational results  \citep{Rea_2011, Francesco_2017}, others on theoretical frameworks \citep{Turolla_2015, tong_2023}, and some offer a synthesis of both \citep{Esposito_2021, Kaspi_2017}. Despite these advances, many magnetar properties remain poorly understood. In particular, the triggering mechanisms of glitches and anti-glitches, as well as the underlying causal connection between glitches and radiative activity, remain elusive. The association of SGR J1935+2154 with FRB 200428 further suggests that at least some fast radio bursts (FRBs)-- transient, extremely energetic, millisecond-duration bursts of radio waves -- may originate from magnetars \citep{Andersen_2020, Bochenek_2020}, opening a new line of inquiry in magnetar astrophysics. Continued, systematic timing observations and the accumulation of glitch and activity samples are crucial for addressing these open questions. The Neutron Star Interior Composition Explorer (\nicer) has provided sustained, high-cadence monitoring of a sample of magnetars over the past eight years. Its excellent time resolution and sensitivity in the soft X-ray band offer unique advantages for investigating the timing and pulse profile properties of magnetars. In this paper, we present a systematic timing analysis of four persistently bright magnetars (\targeta, \targetb, \targetc, and \targetd) based on \nicer observations from 2017 to 2024. We focus on timing properties and pulse profile evolution. Analyses of flux variability and spectral properties will be addressed in a future study.

The structure of this paper is as follows: In Section~\ref{sec:rev}, we provide a brief overview of the four AXPs. Section~\ref{sec:obs} describes the observations and the data reduction procedures. In Section~\ref{sec:ana}, we summarize the analysis methods. Results are presented in Section~\ref{sec:res}, followed by a discussion of findings and conclusions in Section~\ref{sec:dis}.

%%%%%%%%%%%%%%%%%%%%%%%%%%%%%%%%%%%%%%%%%%%%%%%%%%%%%%%%%%%%%%%%%%%%%%%%%%%%%%%%%%%%%%%%%%%%%%%%%
\section{A SHORT REVIEW OF FOUR TARGETS} \label{sec:rev}

\subsection{\targeta} \label{sec:rev-1}

\targeta was the first discovered AXP, with a spin period of approximately 6.8 $\mathrm{s}$, located within the supernova remnant CTB 109 \citep{Fahlman_1981}.

\targeta entered an outburst phase in 2002, during which a glitch occurred, accompanied by changes in pulsed profile, flux, and other properties \citep{Kaspi_2003, Woods_2004}. In 2007, the source experienced a second glitch, which, unlike the first, was in a radiatively quiet state and did not show any changes in the pulsed profile \citep{Dib_2014}. In 2009, a timing anomaly was also observed, accompanied by changes in pulsed flux and the pulsed profile \citep{Icdem_2012}. In 2012, the source underwent its first anti-glitch,detected in \swift observations and accompanied by a short burst \citep{Archibald_2013}.  \citet{Younes_2020} analyzed the observational data from \swift and \nicer on this source from 2013 to 2019, identifying one glitch and one anti-glitch, both occurring in radiatively quiet states.

\subsection{\targetb} \label{sec:rev-2}

\targetb is one of the brightest persistent magnetars, detected by the Uhuru mission \citep{Giacconi_1972}, and later identified as a pulsar with a period of 8.7 $\mathrm{s}$ \citep{Israel_1994}.

During a 16-year monitoring campaign with \rxte, the source entered an active phase in 2006, characterized by several short bursts and significant variations in timing behavior, pulse profile, and flux. \citet{Dib_2014} classified this event as a glitch candidate, while \citet{Gavriil_2011a} suggested that it was a net spin-down glitch. In 2011, the source showed a large glitch, accompanied by a short burst detected by \swift. 
Additionally, \swift observed some short bursts from this source in 2015, during which an over-recovery glitch occurred, which resulted in a net spin-down because of significant recovery behavior following the initial spin-up glitch.

\subsection{\targetc} \label{sec:rev-3}

\targetc (hereafter referred to as \targetcc for brevity) is an 11 $\mathrm{s}$ AXP first discovered as a pulsar by \citet{Sugizaki_1997} by using ASCA data. Compared to \targeta and \targetb, this source exhibits more frequent timing anomalies.

The first glitch was detected in 1999 \citep{Kaspi_2000}, marking the first glitch ever observed in a magnetar -- an expected phenomenon under the magnetar model.  The second glitch was in 2001 with an exponential recovery \citep{Kaspi_2003b}, and the third glitch occurred in 2005 \citep{Dib_2008}. All three events were radiatively quiet, showing no significant changes in pulse profile or flux. \citet{Israel_2007} also reported several glitches, but \citet{Dib_2014} reclassified them as glitch candidates after considering timing noise. Additionally, \citet{Scholz_2014} discovered a new glitch using \rxte and \swift data.

\subsection{\targetd} \label{sec:rev-4}

\targetd is located at the center of the supernova remnant Kes 73, and was first classified as an AXP with a spin period of approximately 11.8 $\mathrm{s}$ \citep{Vasisht_1997}. This source exhibits significant timing noise, with notable variations in its rotational frequency derivative on timescales of about a year. During the \rxte monitoring campaign, four glitches were observed in 2002, 2003, 2006, and 2007. Interestingly, all four glitches were radiatively quiet, showing no significant changes in radiative behavior—such as bursts or pulse profile variations \citep{Dib_2008, Dib_2014}. It is worth noting that an anti-glitch was reported by \citet{Sinem_2014}, although this was not confirmed in the results of \citet{Dib_2014}. In addition, \citet{An_2015} presented timing results from a three-year \swift observation campaign beginning in 2011, which revealed noisy timing behavior and large timing residuals. In August 2024, the source entered an active phase, which is markedly different from the previously quiet state. Almost all radiative behaviors changed, accompanied by a glitch \citep{Younes_2025}.

%%%%%%%%%%%%%%%%%%%%%%%%%%%%%%%%%%%%%%%%%%%%%%%%%%%%%%%%%%%%%%%%%%%%%%%%%%%%%%%%%%%%%%%%%%%%%%%%%
\section{OBSERVATION AND DATA REDUCTION \label{sec:obs}}

\begin{deluxetable*}{ccccc}
\caption{Summary of Spin Properties and X-ray Data}
\label{table:obs-used}
\startdata
\\[1mm]
Parameter & \targeta & \targetb & \targetc & \targetd \\
\hline
Frequency,~$\nu$~(Hz) & 0.1433 & 0.1151 & 0.0908 & 0.0847  \\
$\dot\nu$~($10^{-13} ~Hz ~s^{-1}$) & -0.097 & -0.26 & -1.68 & -2.92  \\
Magnetic field,~ $B$~($10^{14}$~G) & 0.59 & 1.3 & 4.7 & 7.0  \\
\hline
Number of Analyzed \nicer Observations              & 132           & 128           & 103           & 119  \\
Total Exposure Time of \nicer Observations~(ks)     & 115.567       & 136.036       & 61.589        & 242.038  \\
Date of the First Analyzed Observations             & 2018 Jan 17   & 2019 Mar 16   & 2019 Mar 14   & 2019 Mar 19  \\
Date of the Last Analyzed Observations              & 2024 Oct 30   & 2024 Nov 13   & 2024 May 17   & 2024 Jun 26  \\
Band Used for the Timing Analysis~(keV)             & 0.8-8         & 0.7-10        & 0.7-10        & 0.7-10  \\
\hline
Number of Analyzed \swift Observations              & 20            & -            & -              & -  \\
Total Exposure Time of \swift Observations~(ks)     & 69.403        & -            & -              & -  \\
Date of the First Analyzed Observations             & 2017 Dec 7    & -            & -              & -  \\
Date of the Last Analyzed Observations              & 2019 Jan 7    & -            & -              & -  \\
Band Used for the Timing Analysis~(keV)             & 0.5-10        & -            & -              & -  \\
\hline
The Observation ID of \ixpe                         & 02007899      & -            & 01003199       & -  \\
The Exposure Time of \ixpe Observations~(ks)        & 328.530       & -            & 837.633        & -  \\
\enddata
\tablecomments{(a) The basic characteristics of the four sources are obtained from the McGill Magnetar Catalog webpage. (b) In this work, only two observations of \ixpe are utilized; therefore, we simply list the observation ID and exposure time.}
\end{deluxetable*}

The Neutron Star Interior Composition Explorer (\nicer) is a payload onboard the International Space Station, dedicated to studying neutron stars through soft X-ray (0.2--12 keV) timing observations \citep{Gendreau_2016}. The \target have been regularly monitored since the mission’s launch on 2017 June 3, with increased cadence after 2019. In this paper, we analyze all data from \nicer observations of four \target up to 2024. We use \texttt{HEASOFT} v6.33.2 to process the \nicer data, initially employing the default standard procedure of \texttt{nicerl2} (part of \texttt{NICERDAS} version v12) to generate cleaned and calibrated event files. We then apply \texttt{nicerl3-lc} to extract light curves in the 12--15 keV band, and use \texttt{maketime} to filter out time intervals with count rates exceeding 0.2 counts s$^{-1}$, which are typically associated with particle-induced flaring. The resulting good time interval (GTI) file is then used to re-run \texttt{nicerl2} to produce the final event file. Barycentric corrections are subsequently applied using the \texttt{barycorr} task. To generate the background spectra, which are later used in the pulse profile analysis, we use \texttt{nicerl3-spect}, selecting the background model \textbf{3C50}. For a more detailed description of the \nicer data processing pipeline, please refer to the webpage \footnote{\url{https://heasarc.gsfc.nasa.gov/docs/nicer/analysis_threads/}}.

After completing the data reduction, we select only those observations with exposure times longer than 50 seconds for timing analysis. Table~\ref{table:obs-used} summarizes the basic properties of the sources, including the total number of \nicer observations analyzed for each target, the start and end dates of the observations, and the energy bands used in the analysis.

Due to the limited number of \nicer observations preceding the anti-glitch event in \targeta, we also include 20 archival \swift observations obtained before the event to better constrain the spin parameters (see Table~\ref{table:obs-used}).  The cleaned event files are rebuilt using the task \texttt{xrtpipeline}, and only the Windowed Timing mode data with high temporal resolution (1.8 ms) are involved in this analysis. The source photons in the 0.5–-10 keV band are extracted from a circular region with a radius of 20 pixels centered at the source position, and then barycentric correction is performed using the task \texttt{barycorr}.

Similarly, to increase our data statistics and improve the accuracy of timing -- particularly in determining the state transition events of \targetc -— we add two \ixpe observations to our timing analysis for both \targeta and \targetc. We perform the data cleaning, region selection, and analysis using the software suite \texttt{ixpeobssim}, which is equipped with various tools to process level 2 event files and produce scientific results \citep{Baldini_2022}. A circular region with a radius of 60" was selected for the extraction of source events, while an annular region with inner and outer radii of 180" and 240" was selected for background events. To improve the signal-to-noise ratio, we combined the data from all three detector units for the timing analysis. For a more detailed data processing pipeline regarding \ixpe, readers can refer to these articles \citep{Peng_2024,Stewart_2025,Rigoselli_2025}. As before, we use task \texttt{barycorr} to perform barycentric correction.

%%%%%%%%%%%%%%%%%%%%%%%%%%%%%%%%%%%%%%%%%%%%%%%%%%%%%%%%%%%%%%%%%%%%%%%%%%%%%%%%%%%%%%%%%%%%%%%%%
\section{ANALYSIS} \label{sec:ana}

\subsection{Timing Analysis \label{sec:ana-1}}

% \subsubsection{$\chi^{2}$ searching and TOA calculation} \label{sec:ana-1-1}
\subsubsection[\texorpdfstring{Chi-squared searching and TOA calculation}{Chi-squared searching and TOA calculation}]{\texorpdfstring{$\chi^{2}$}{Chi-squared} searching and TOA calculation} \label{sec:ana-1-1}

We begin the timing analysis with a $\chi^2$ search, identifying the spin frequency that yields the largest deviation from a uniform distribution in the folded pulse profile—quantified by the maximum $\chi^2$ value -—as the best-fit spin frequency for a barycenter-corrected, binned time series \citep{Ge_2012}. This search is performed approximately every 20 observations, with the exact interval adjusted based on the observation cadence and noise level. A preliminary ephemeris is constructed for each subset, and the corresponding pulse profiles are aligned and summed to generate a template pulse profile.

After obtaining the template pulse profile, we compute the phase shift $\phi_{0}$ between each individual pulse profile and the template by means of cross-correlation \citep{zhou_2022b}. The corresponding time of arrival (TOA) is then calculated using the following equation:
\begin{equation}
\text{TOA} = T_{0} + \frac{\phi_{0}}{\nu}.
\label{eq01}
\end{equation}
where $T_{0}$ denotes the time of the reference epoch for each observation (typically the starting point or midpoint) and $\nu$ is the spin frequency at that epoch. To estimate the uncertainty associated with each TOA, we introduce Poisson noise to each pulse profile to simulate 3000 noisy profiles, which are subsequently cross-correlated with the template. The uncertainty of the phase is ultimately determined by the standard deviation of a series of returned values from the cross-correlation \citep{Dib_2014}.

\subsubsection{Phase-coherent timing analysis} \label{sec:ana-1-2}

After obtaining the TOAs, we utilize the TEMPO2 software for timing analysis, which returns the spin parameters and residuals \citep{Hobbs_2006}. We perform timing analysis in two different ways for four AXPs: partially phase-coherent timing (PPCT) and fully phase-coherent timing (FPCT). The former divides the data set into relatively short subsets and performs timing analysis by individually phase-connecting these subsets. This approach can reduce the effects of timing noise and covariances caused by polynomial fitting when determining the spin parameters, particularly when the data set is long enough to necessitate the use of many higher-order frequency derivatives. In contrast, FPCT phase-connects the entire data set, which enables us to confirm the results obtained from PPCT and obtain a more precise timing solution, especially regarding the parameter values of glitches \citep{Livingstone_2005}.

Young pulsars--especially magnetars--often exhibit substantial timing noise,  which can affect the measurement of intrinsic frequency parameters as well as the identification of glitches. To mitigate the impact of such noise on long-term timing evolution, PPCT is applied to segmented time intervals. By analyzing multiple shorter time spans, we can more clearly trace the evolution of spin parameters \citep{Ferdman_2015}. When no glitches are present, the long-term evolution of the TOAs in the phase domain can be modeled using a Taylor expansion:
\begin{equation}
\phi(t)=\phi_0+\nu{(t-t_{0})}+\frac{1}{2}\dot\nu{(t-t_{0})^{2}}+\frac{1}{6}\ddot\nu{(t-t_{0})^{3}}+... ,
\label{eq02}
\end{equation}
where $\nu$, $\dot\nu$ and $\ddot\nu$ are the frequency, frequency derivative and the second derivative of frequency of the reference time $t_{0}$, respectively \citep{Wong_2001}.

We typically fit the timing model over intervals of approximately 100 days, with a $\sim$ 50-day overlap between segments. These durations are also adjusted based on the level of timing noise and the TOA density. For most segments, we include only the spin frequency $\nu$ and its derivative $\dot{\nu}$ in the fit, except in cases of particularly strong timing noise. The reference epoch is chosen at the midpoint of each time segment \citep{Dib_2014}. If a glitch occurs, it manifests as a discontinuity in the residuals. PPCT segments are therefore chosen to avoid spanning glitch epochs.

Following the initial PPCT results, we apply FPCT to derive more accurate and continuous timing solutions \citep{GE_2022, Younes_2020}. When a glitch occurs, Equation~\ref{eq02} is no longer valid, and the residuals deviate significantly. In such cases, FPCT assumes that the phase residual model conforms to Equation:
\begin{equation}
\begin{aligned}
\Delta\phi(t) & = \Delta\nu\delta{t} + \frac{1}{2}\Delta\dot\nu{\delta{t}^{2}} + \Delta\nu_{d}\tau\left[1-e^{-\frac{\delta t}{\tau}}\right] \\
\delta{t} & = t - t_{g} > 0
\end{aligned}
\label{eq03}
\end{equation}
where \( t_g \) is the time when the glitch occurs, \( \Delta \nu \) and \( \Delta \dot{\nu} \) are the changes in frequency and frequency derivative, respectively. Sometimes, an exponential decay term needs to be added; \( \tau \) is the duration for which the suddenly changed frequency \( \Delta \nu_d \) returns to 0 \citep{Tuo_2024, Wong_2001}. For each identified glitch, we perform FPCT using time intervals sufficiently long to capture the behavior both before and after the event, allowing us to determine refined timing parameters that account for both permanent and transient changes.

\subsection{Pulse profile Analysis} \label{sec:ana-2}

For each source, we divide the observations into separate time segments based on the epochs of detected glitches, except for \targetd, where we used the observation gaps as segmentation points. For each segment, the corresponding ephemeris is used to fold the pulse profiles of all included observations. During this process, we calculate phase offsets between ephemerides to ensure that the resulting folded pulse profiles are phase-aligned across segments. To extract the net pulse profiles, we first estimate the background photon counts for each observation using its background spectrum. The background level is averaged over the phase intervals of each pulse profile, assuming that it remains stable within the short (few-second) exposure durations. Subtracting the estimated background yields the net pulse profile for each observation. We then sum the net pulse profiles from different time segments to obtain the net average pulse profiles. To compare profiles before and after a glitch, all pulse profiles are normalized by dividing by their respective mean count rates.

\section{RESULTS} \label{sec:res}

\subsection{\targeta} \label{sec:res-1}

\begin{figure*}
\begin{center}
\includegraphics[width=0.65\textwidth,clip]{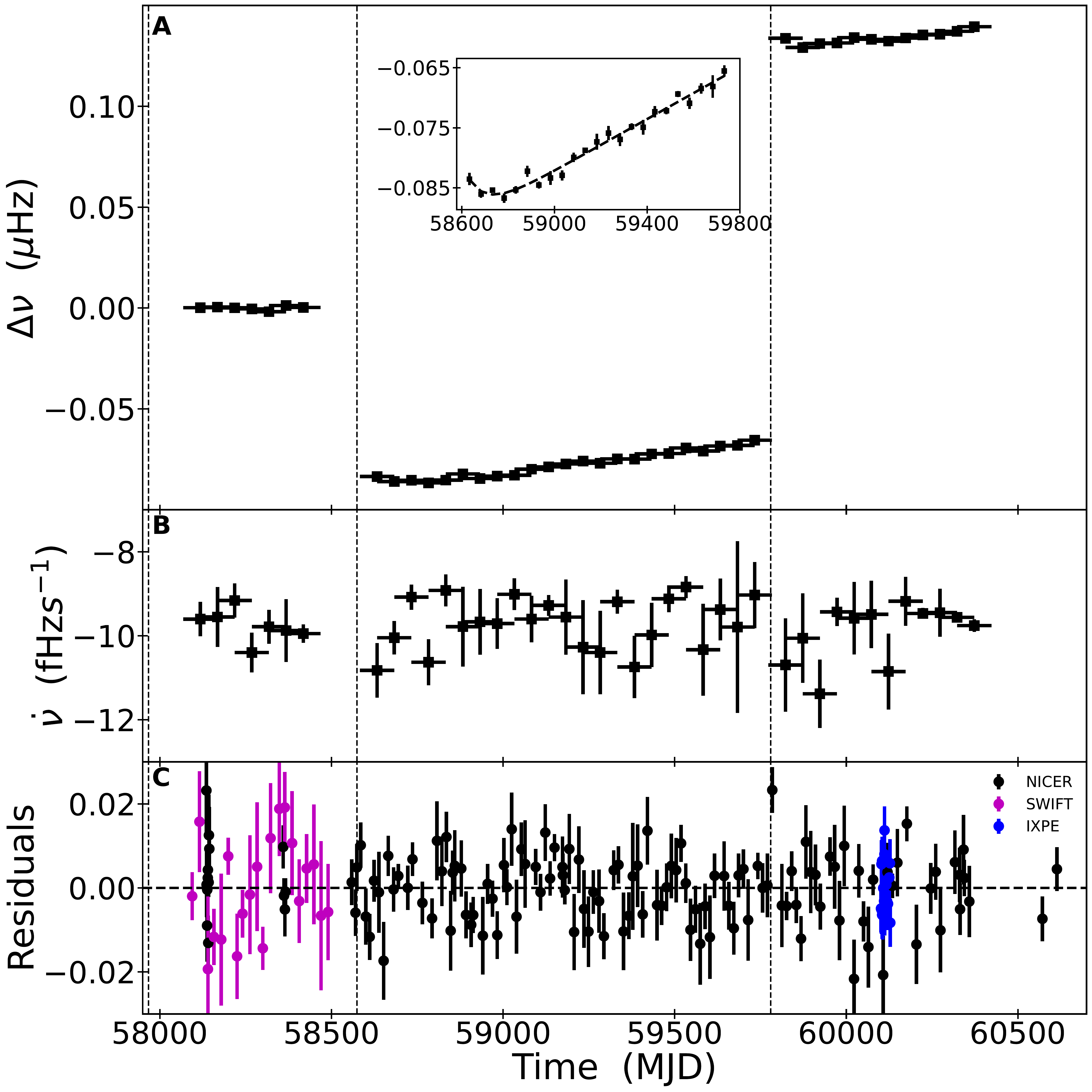}
\caption{The long-term timing evolution of \targeta. Three vertical dashed lines represent the times of the glitches within the \nicer observation range. Panel(A), Frequency as a function of time with a linear trend in frequency subtracted. The inset graph zooms in on the results of the anti-glitch, revealing the existence of an exponential term. Panel(B), Frequency derivative as a function of time. Panel(C), The timing residuals of all TOAs, black, magenta, and blue data represent from \nicer, \swift and \ixpe , respectively.} 
\label{timing:a}
\end{center}
\end{figure*}

\begin{figure*}
\begin{center}
\includegraphics[width=0.75\textwidth,clip]{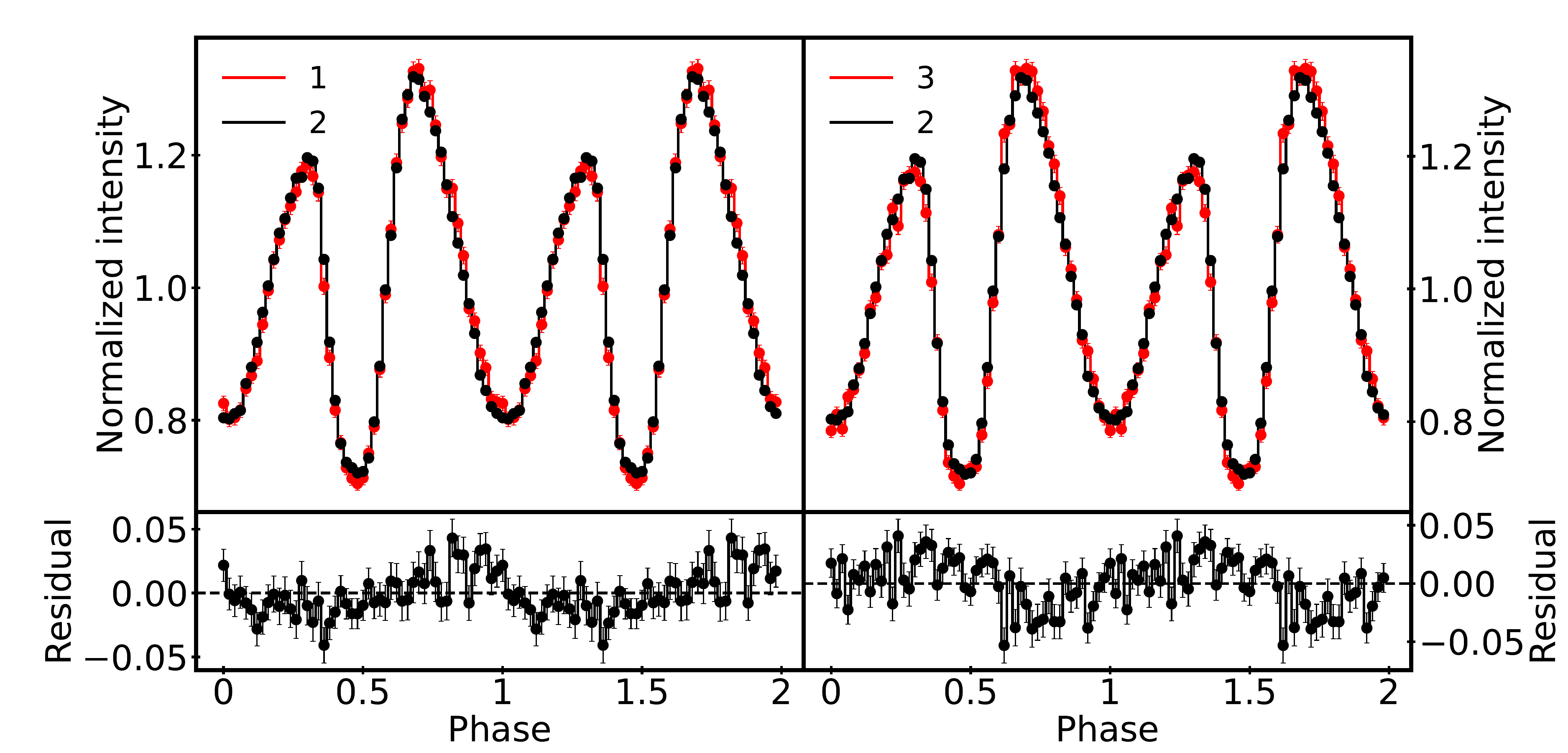}
\caption{Comparison results of normalized pulse profiles of \targeta . The numbers 1-3 represent the average pulse profiles for the three time segments, divided based on the glitch events shown in Figure \ref{timing:a} (with the first glitch excluded). The residuals for each pair of normalized pulse profiles, obtained by subtracting one profile from the other, are plotted below the corresponding panels.}
\label{pro:a}
\end{center}
\end{figure*}

A summary timing behavior of \targeta as observed by \nicer is shown in Figure \ref{timing:a}. The long-term timing parameters of the source are listed in Table \ref{table:a} and the pulse profiles are shown in Figure \ref{pro:a}.

Panels (a) and (b) of figure \ref{timing:a} show that \targeta exhibits low timing noise compared to other AXPs, which is consistent with previous findings \citep{Dib_2014}. As indicated by the black vertical dashed lines in figure \ref{timing:a}, three large glitches occurred in \targeta during the time span of our analysis.

\begin{deluxetable}{lc}
\tablewidth{0pt}
\label{table:a}
\tablecaption{Spin parameters of \targeta with glitches}
\startdata
  &  \\
Parameters            & Value    \\
\hline
R.A.                                                            & 23:01:08.295  \\
Decl.                                                           & +58:52:44.45  \\
MJD range                                                      & 58029.4-60672.0 \\
Epoch (MJD)                                                    & 58300  \\
$\nu $(${\rm Hz}$)                                             & 0.1432825964(1)  \\
$\dot\nu$(${\rm fHz\cdot s^{-1}}$)                             & -9.77(2)    \\
\hline
$t_{\mathrm{g1}}$(MJD)                                                  &58575  \\
$\triangle\phi_{\mathrm{g1}}$                                           &-0.012(7)  \\
$\triangle\nu_{\mathrm{g1}}$($\rm 10^{-8} ~Hz$)                           &-8.95(5)  \\
$\triangle\dot\nu_{\mathrm{g1}}$($\rm 10^{-16} ~Hz ~s^{-1}$)          &3.2(1)  \\
$\triangle\nu_{\mathrm{d1}}$($\rm 10^{-8} ~Hz$)                           &1.2(2)  \\
$\tau_{\mathrm{d1}}$(day)                                               &89(17)  \\
$\triangle\nu_{\mathrm{g1}}/\nu_{\mathrm{g1}}$($10^{-7}$)                        &-6.25(3)  \\
$\triangle\dot\nu_{\mathrm{g1}}/\dot\nu_{\mathrm{g1}}$                           &-0.033(1)  \\
\hline
$t_{\mathrm{g2}}$(MJD)                                                  &59780  \\
$\triangle\phi_{\mathrm{g2}}$                                           &0.066(4)  \\
$\triangle\nu_{\mathrm{g2}}$($\rm 10^{-7} ~Hz$)                           &1.938(2)  \\
$\triangle\dot\nu_{\mathrm{g2}}$($\rm 10^{-17} ~Hz ~s^{-1}$)           &-8.9(9)  \\
$\triangle\nu_{\mathrm{g2}}/\nu_{\mathrm{g2}}$($10^{-6}$)                        &1.352(2)  \\
$\triangle\dot\nu_{\mathrm{g2}}/\dot\nu_{\mathrm{g2}}$                           &0.009(1)  \\
Residuals (ms)                                                 &48.554     \\
\enddata
\tablecomments{The subscripts g1 and g2 represent the parameters of the first and second glitch, respectively. The subscript d denotes the exponential decay term. The uncertainties of all the solutions to the spin parameters obtained using tempo2 in this paper are at the default confidence level of $1\sigma$}.
\end{deluxetable}

\begin{figure*}
\begin{center}
\includegraphics[width=0.65\textwidth,clip]{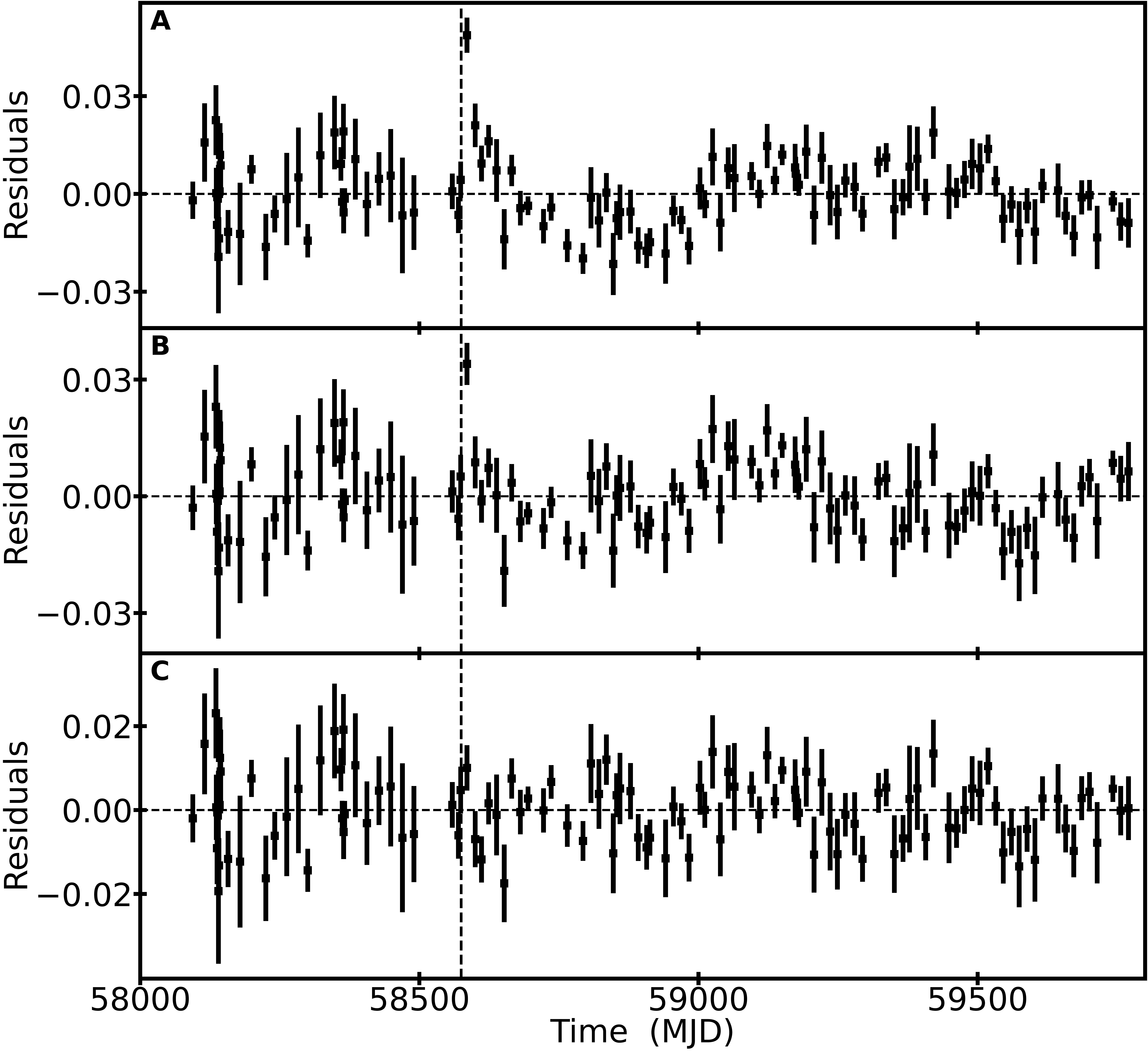}
\caption{The timing residuals of the anti-glitch for \targeta obtained from three different models. Panel(A), excluding the exponential term, the model utilized only $\nu$, $\dot{\nu}$, $\Delta \nu$, $\Delta \dot{\nu}$, with an RMS residual of 68.988 ms. Panel(B), the second derivative of frequency $\ddot{\nu}$ was added to the analysis from Panel(A), with an RMS residual of 57.448 ms. Panel(C), the exponential term $\Delta\nu_{d}$ was added to the analysis from Panel(A), with an RMS residual of 46.865 ms.}
\label{residual-a}
\end{center}
\end{figure*}

A glitch event occurred in \targeta shortly after the launch of \nicer. Since only two data points are available before this glitch, we mark the time of the event with a black vertical line and treat it as the starting point of our timing analysis. Further details about this glitch can be found in \citet{Younes_2020}. The second timing event is an anti-glitch, which is also reported by  \citet{Younes_2020}. However, we find that this anti-glitch is accompanied by an exponential decay phenomenon, which was not identified before.  If this feature is not included, the subsequent TOAs cannot be well fitted, as shown in Figure \ref{residual-a}. This discrepancy arises because their dataset covers only about 100 days post-event, insufficient to reveal the recovery trend. Indeed, when we use the same time span as in their study, we reproduce their results. In this work, we use a sufficiently longer dataset and derive a more accurate spin parameters, as shown in Table \ref{table:a}. The timescale of the exponential recovery associated with this anti-glitch is estimated to be $89 \pm 17$ days.
This exponential timescale can be explained by the creep response of vortex lines against flux tubes in the interior of a neutron star heated by magnetar activity, provided that the toroidal field in the core is a fraction of the polar magnetic field and the magnetar cools down by the direct Urca process \citep{Erbil_2017a}. Furthermore, the increase in temperature due to magnetar burst activity can perturb the equilibrium angular velocity lag between the crust and the superfluid, leading to inward vortex motion and in turn anti-glitch \citep{Erbil_2016}. Such a deviation from the steady-state vortex line distribution would produce a stronger radially outward vortex current, which is expected to lead to a large spin-up glitch via excess angular momentum transfer from the superfluid to the crust \citep{Erbil_2019}. Indeed, following the anti-glitch, we detect a new glitch.

% Following the anti-glitch, we detect a new glitch.
Owing to the low timing noise, a straightforward ephemeris is able to fit all TOAs covering a 7-year span, as shown in Table \ref{table:a}. Panel (c) of figure \ref{timing:a} displays the residuals after subtracting our best-fit model.

\begin{deluxetable}{lccc}
\tablewidth{0pt}
\label{table:a-s}
\tablecaption{Spin parameters of \targeta without glitches}
\startdata
  &  \\
Parameters            & R.A.                 & Decl.    \\
Value                 & 23:01:08.295        & +58:52:44.45   \\
\hline
Parameters            & S1           & S2    & S3   \\
\hline
MJD range                                             & 58068-58584      & 58584-59783      & 59783-60677 \\
Epoch (MJD)                                           & 58300            & 59000            & 60000  \\
$\phi_0$                                              & 0.832(2)           & 0.2612(7)           & 0.205(1) \\
$\nu $(${\rm Hz}$)                                    & 0.1432825964(1)  & 0.14328192855(6)  & 0.1432813039(1)  \\
$\dot\nu$(${\rm fHz\cdot s^{-1}}$)                    & -9.77(1)         & -9.467(3)         & -9.534(6)    \\
Residuals (ms)                                        & 52.397           & 74.030           & 51.778     \\
\enddata
\tablecomments{S1, S2, and S3 represent the three time ranges obtained by using the glitch epochs as segmentation points.}
\end{deluxetable}

We divide the data into three segments based on the epochs of the glitches (excluding the two early observations prior to the first glitch), and construct an average pulse profile for each segment. Each profile consists of 50 phase bins based on the corresponding ephemeris shown in Table \ref{table:a-s}. We normalize and compare the pulse profiles before and after each glitch, as shown in Figure \ref{pro:a}. The pulse profile of \targeta is characterized by two peaks of varying heights, consistent with previous studies. 
We find that although there are no significant characteristic changes (such as the transition from single-peaked to multi-peaked profiles observed before and after the \targetd burst, as described below), the pulse profiles accumulated over a longer period do exhibit changes, as reflected in the statistical values from the chi-squared test (Table \ref{table:P_value}). This is consistent with the previous idea that magnetars exhibit slow, low-level changes, which can only be detected by summing the pulse profiles over an extended period of time \citep{Dib_2008, Dib_2014}.

\subsection{\targetb} \label{sec:res-2}
\begin{figure*}
\begin{center}
\includegraphics[width=0.65\textwidth,clip]{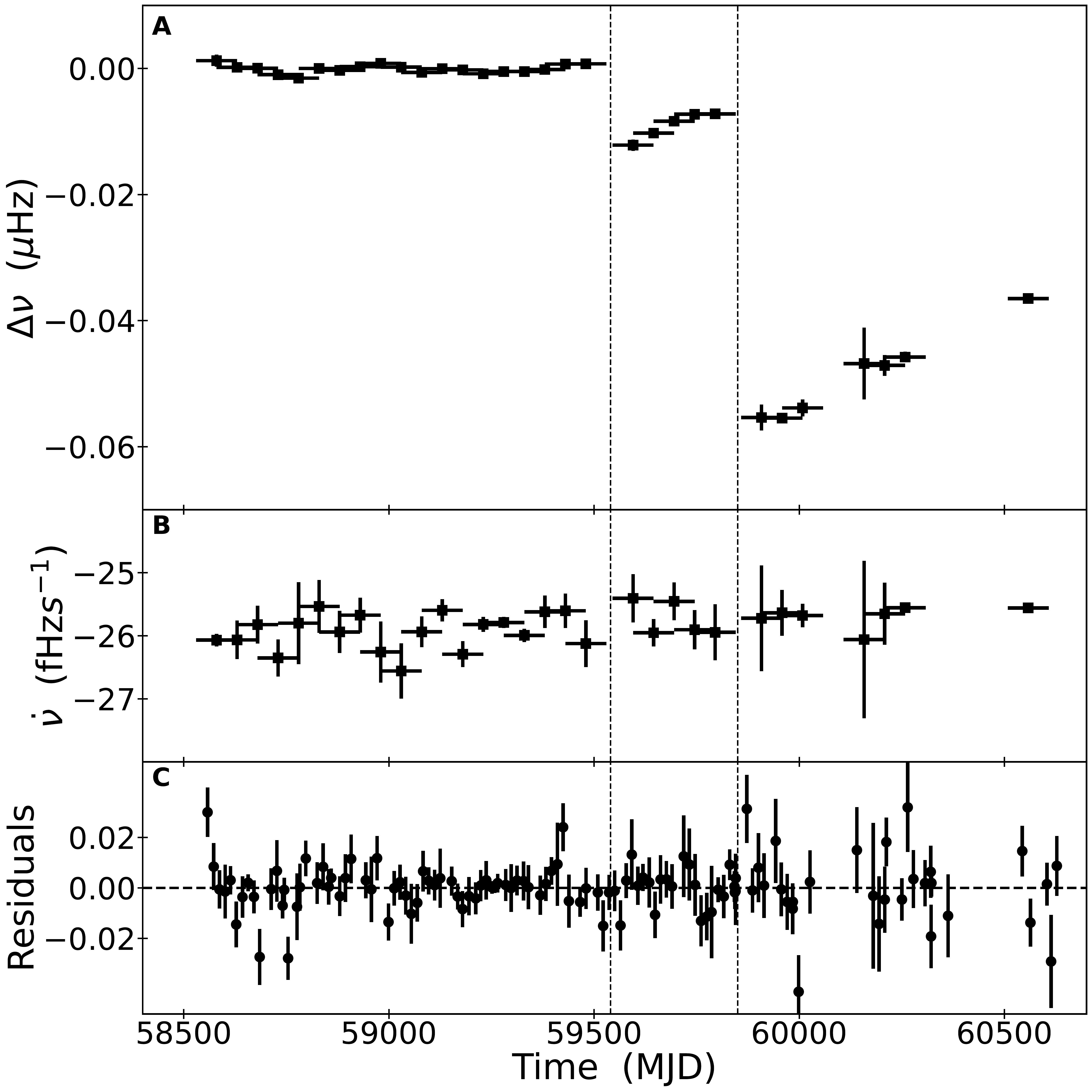}
\caption{The long-term timing evolution of \targetb. Two black vertical dashed lines represent the times of the glitches within the \nicer observation range. Panel(A), Frequency as a function of time with a linear trend in frequency subtracted. Panel(B), Frequency derivative as a function of time. Panel(C), The timing residuals of all TOAs.} 
\label{timing:b}
\end{center}
\end{figure*}

\begin{figure*}
\begin{center}
\includegraphics[width=0.75\textwidth,clip]{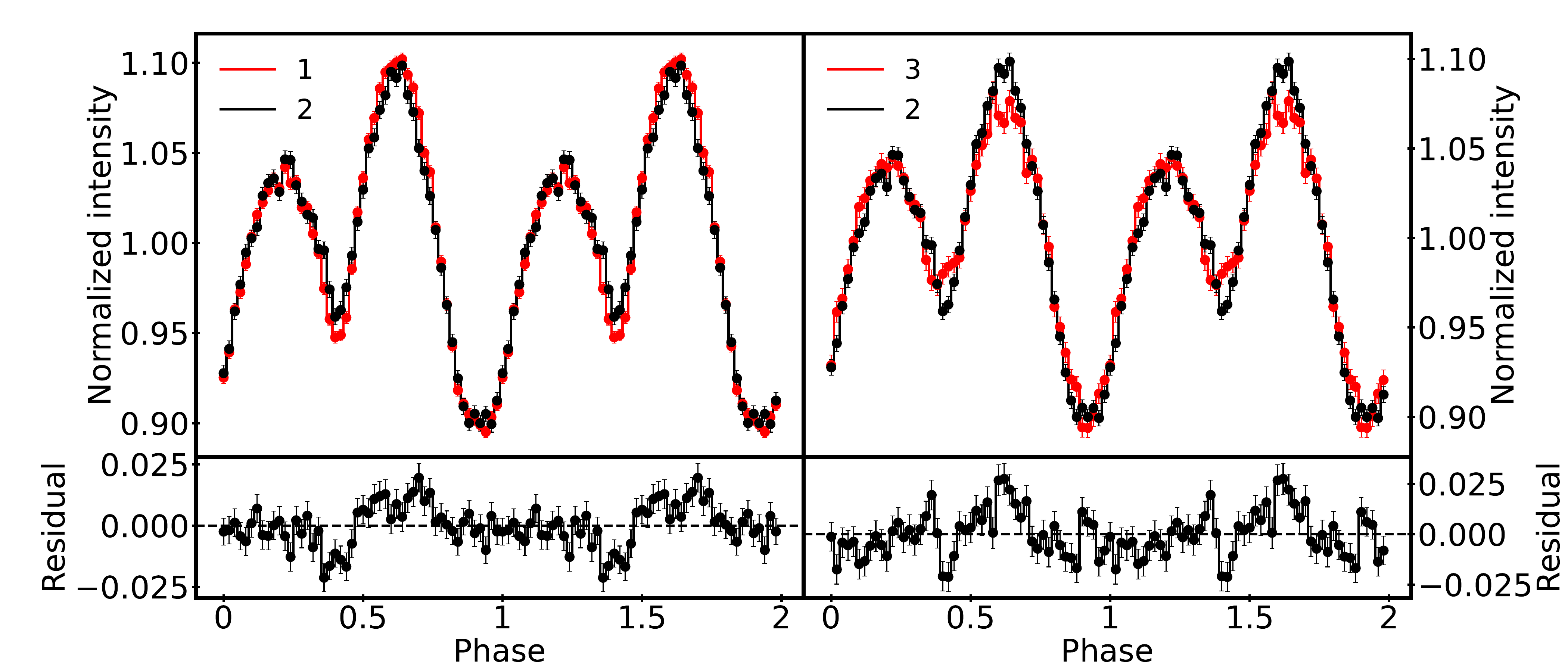}
\caption{Comparison results of normalized pulse profiles of \targetb . The numbers 1-3 represent the average pulse profiles for the three time segments, divided based on the glitch events shown in Figure \ref{timing:b}. The residuals for each pair of normalized pulse profiles, obtained by subtracting one profile from the other, are plotted below the corresponding panels.} 
\label{pro:b}
\end{center}
\end{figure*}

A summary of the behavior of \targetb as observed by \nicer is shown in Figure \ref{timing:b}. The long-term timing parameters of the source are presented in Table \ref{table:b} and the pulse profiles are shown in Figure \ref{pro:b}.

As can be seen in figure \ref{timing:b}, two notable  timing events occurred during the time span of our analysis, marked by the vertical dashed lines.

\begin{deluxetable}{lc}
\tablewidth{0pt}
\label{table:b}
\tablecaption{Spin parameters of \targetb with glitches}
\startdata
  &  \\
Parameters                                                     & Value    \\
\hline
R.A.                                                            & 01:46:22.21  \\
Decl.                                                           & +61:45:03.8  \\
MJD range                                                      & 58504.0-60730.1 \\
Epoch (MJD)                                                    & 59000  \\
$\nu $(${\rm Hz}$)                                             & 0.11508073213(4)  \\
$\dot\nu$(${\rm 10^{-14} ~Hz ~s^{-1}}$)                        & -2.5914(3)    \\
\hline
$t_{\mathrm{g1}}$(MJD)                                                  &59540  \\
$\triangle\phi_{\mathrm{g1}}$                                           &-0.002(6)  \\
$\triangle\nu_{\mathrm{g1}}$($\rm 10^{-8} ~Hz$)                           &-1.25(9)  \\
$\triangle\dot\nu_{\mathrm{g1}}$($\rm 10^{-16} ~Hz ~s^{-1}$)          &2.8(6)  \\
$\triangle\nu_{\mathrm{g1}}/\nu_{\mathrm{g1}}$($10^{-7}$)                        &-1.09(9)  \\
$\triangle\dot\nu_{\mathrm{g1}}/\dot\nu_{\mathrm{g1}}$                           &-0.011(2)  \\
\hline
$t_{\mathrm{g2}}$(MJD)                                                  &59850  \\
$\triangle\phi_{\mathrm{g2}}$                                           &-0.044(6)  \\
$\triangle\nu_{\mathrm{g2}}$($\rm 10^{-8} ~Hz$)                           &-5.42(9)  \\
$\triangle\dot\nu_{\mathrm{g2}}$($\rm 10^{-16} ~Hz ~s^{-1}$)          &1.0(7)  \\
$\triangle\nu_{\mathrm{g2}}/\nu_{\mathrm{g2}}$($10^{-7}$)                        &-4.71(8)  \\
$\triangle\dot\nu_{\mathrm{g2}}/\dot\nu_{\mathrm{g2}}$                           &-0.004(3)  \\
Residuals (ms)                                                 &61.515        \\
\enddata
\tablecomments{The subscripts g1 and g2 represent the parameters of the first and second glitch, respectively.}
\end{deluxetable}

\begin{deluxetable}{lccc}
\tablewidth{0pt}
\label{table:b-s}
\tablecaption{Spin parameters of \targetb without glitches}
\startdata
  &  \\
Parameters            & R.A.                 & Decl.    \\
Value                 & 01:46:22.21         & +61:45:03.8   \\
\hline
Parameters            & S1           & S2    & S3   \\
\hline
MJD range                                             & 58500-59546      & 59546-59860      & 59860-60700 \\
Epoch (MJD)                                           & 59000            & 59700            & 60200  \\
$\phi_0$                                              & 0.054(1)           & 0.918(2)           & 0.843(3) \\
$\nu $(${\rm Hz}$)                                    & 0.11508073213(3)  & 0.1150791561(1)  & 0.1150779975(1)  \\
$\dot\nu$(${\rm 10^{-14} ~Hz ~s^{-1}}$)               & -2.5914(3)         & -2.564(5)         & -2.554(1)    \\
Residuals (ms)                                        & 54.496           & 50.639           & 104.110     \\
\enddata
\tablecomments{S1, S2, and S3 represent the three time ranges obtained by using the glitch epochs as segmentation points.}
\end{deluxetable}

Over the seven-year \nicer monitoring campaign, \targetb exhibited two timing anomaly events occurring at  closely spaced epochs (59540 and 59850 MJD), suggesting that glitch events in this source do not follow a regular periodic pattern. Interestingly, both timing anomalies are anti-glitches, with $\triangle\nu_{g1} = -1.25(9) \times 10^{-8}$ Hz and $\triangle\nu_{g2} = -5.42(9) \times 10^{-8}$ Hz . The magnitudes of the frequency change are consistent with earlier observations. \citet{Archibald_2017} previously reported two similar spin-down glitches in this source. Including these two spin-down glitch events, source \targetb has become the one with the highest cyan rate of such occurrences; however, the underlying cause of this phenomenon remains unknown. Probably the strong toroidal magnetic field within the crust is evolving in a way that can intermittently release magnetic stress, thereby facilitating vortex-mediated transfer of angular momentum from the magnetar's crust to its interior \citep{Thompson_2000}. 
Such stress-release episodes may be radiatively loud, producing bursts/outbursts, but they may also be radiatively quiet and leave little or no detectable radiative signature.  During the \nicer observation period, the timing behavior of \targetb is relatively stable, enabling the use of a single timing solution to fit all TOAs, as listed in Table \ref{table:b}.

We divide the data into three intervals based on the epochs of the glitches and construct an average pulse profile for each segment.Each profile consists of 50 phase bins based on the corresponding ephemeris shown in Table \ref{table:b-s}. We normalize and compare the pulse profiles before and after each glitch, as shown in Figure \ref{pro:b}. The pulse profile of \targetb is also characterized by two peaks of varying heights, consistent with previous results. Our analysis reveals that the pulse profile of source \targetb demonstrates a regular evolutionary trend, marked by an upward shift in the valley between the secondary peak and the main peak (a finding consistently confirmed by segmenting the data into additional time intervals). Notably, the main peak exhibits significant changes before and after the second glitch event, whereas no such changes are observed around the first glitch. This suggests that the observed variation is not a result of regular evolution.
The chi-squared test values (Table \ref{table:P_value}) also indicate that there are changes in the pulse profiles. This is consistent with the previous idea that magnetars exhibit slow, low-level changes, which can only be detected by summing the pulse profiles over an extended period of time \citep{Dib_2008, Dib_2014}.

\centerwidetable
\begin{deluxetable*}{lccccc}
\tabletypesize{\footnotesize}
\tablewidth{0pt}
\label{table:c}
\tablecaption{Spin parameters of \targetc with glitches}
\startdata
  &  &  \\
Parameters        & R.A.                & Decl.         \\
Value             & 17:08:46.87        & -40:08:52.44  \\
\hline
Parameters  & S1    & S2    & S3     & S4    & mode-switching \\
\hline
MJD range   & 58550-59290   & 58760-59535    & 59290-59760    & 59870-60510    & 59760-59870  \\
Epoch (MJD)                                                    & 58650      & 59000      & 59400      & 60100      & 59800  \\
$\nu $(${\rm Hz}$)                                             & 0.0908133608(5)    & 0.0908084107(3)    & 0.0908026601(5)    & 0.0907921232(4)    & 0.090796633(2)    \\
$\dot\nu$(${\rm \times 10^{-13} ~Hz ~s^{-1}}$)               & -1.665(2)    & -1.6881(2)    & -1.694(2)    & -1.6621(8)    & -2.27(3)    \\
$\ddot\nu$(${\rm \times 10^{-23} ~Hz ~s^{-2}}$)           & 2.0(5)    & 1.9(5)    & -    & -    & 1025(151)    \\
\hline
Parameters  & G1    & G2    & G3     & G4    & mode-switching \\
\hline
$t_{g}$(MJD)                                                   & 58755      & 59290      & 59540      & 60200    & - \\
$\triangle\phi_{g}$                                            & 0.077(9)   & 0.040(9)   & -0.01(3)   & 0.08(2)  & - \\
$\triangle\nu_{g}$($\rm \times 10^{-8} ~Hz$)                            & 13.7(2)    & 8.4(2)     & 2.9(6)     & 2.2(3)  & - \\
$\triangle\dot\nu_{g}$($\rm \times 10^{-15} ~Hz ~s^{-1}$)           & 2.9(3)     & 1.3(2)     & 1.4(5)     & 1.4(2)    & - \\
$\triangle\nu_{g}/\nu_{g}$($\times 10^{-7}$)                          & 15.1(2)    & 9.2(2)     & 3.2(7)     & 2.5(3)  & - \\
$\triangle\dot\nu_{g}/\dot\nu_{g}$                             & 0.018(2)   & 0.007(1)   & -0.008(3)  & 0.008(1)  & - \\
Residuals (ms)                                                 & 96.630      & 98.523     & 118.789    & 80.238      & 53.456        \\
\enddata
\tablecomments{S1 to S4 represent the four time ranges that include G1 to G4.}
\end{deluxetable*}

\subsection{\targetc} \label{sec:res-3}

\begin{figure*}
\begin{center}
\includegraphics[width=0.60\textwidth,clip]{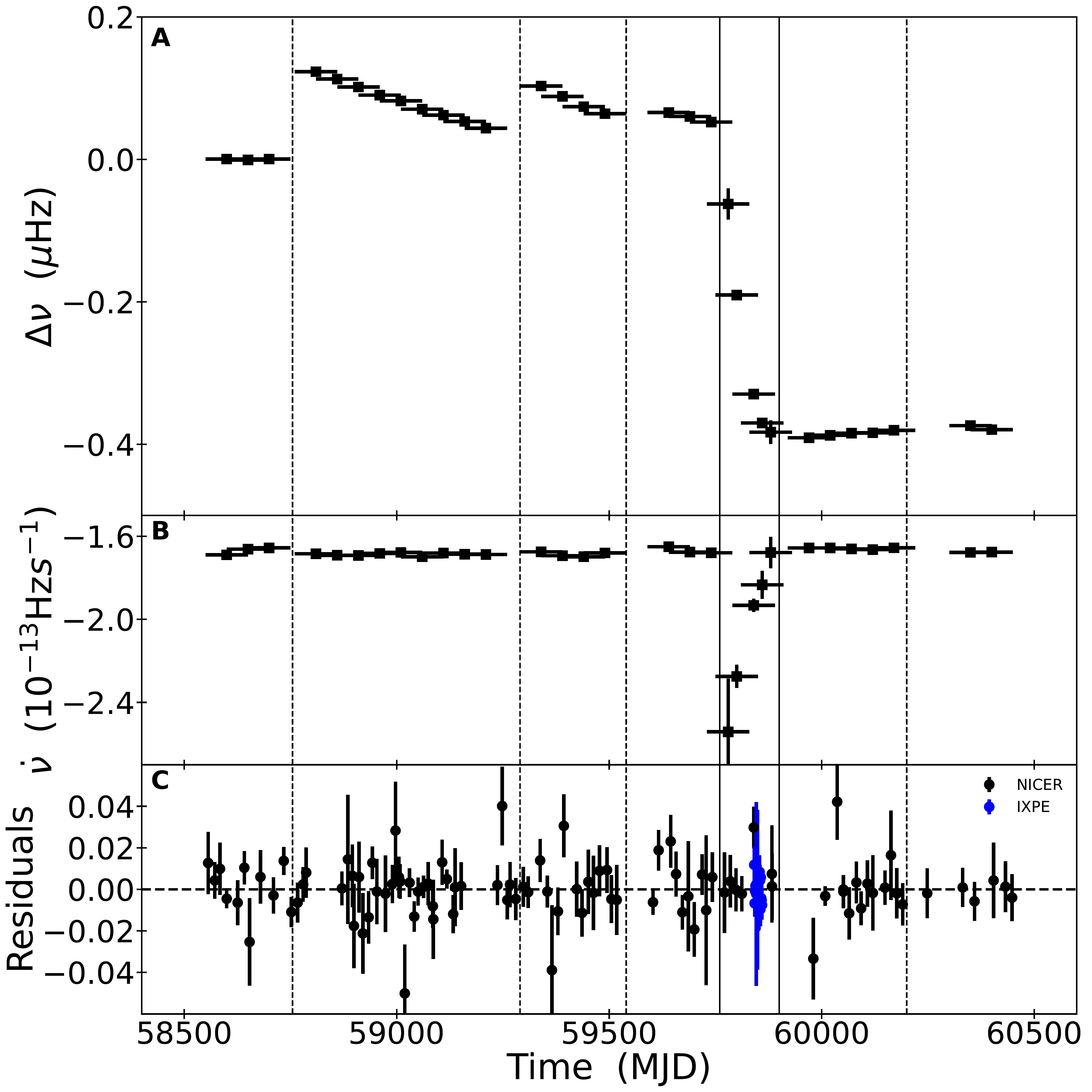}
\caption{The long-term timing evolution of \targetcc. Four vertical dashed lines represent the times of the glitches within the \nicer observation range. Between the two vertical solid lines, a state transition scenario occurred. Panel(A), Frequency as a function of time with a linear trend in frequency subtracted. Panel(B), Frequency derivative as a function of time. Panel(C), The timing residuals of all TOAs, black and blue data represent from \nicer and \ixpe , respectively.} 
\label{timing:c}
\end{center}
\end{figure*}

\begin{figure*} 
\centering
\includegraphics[width=1.0\textwidth]{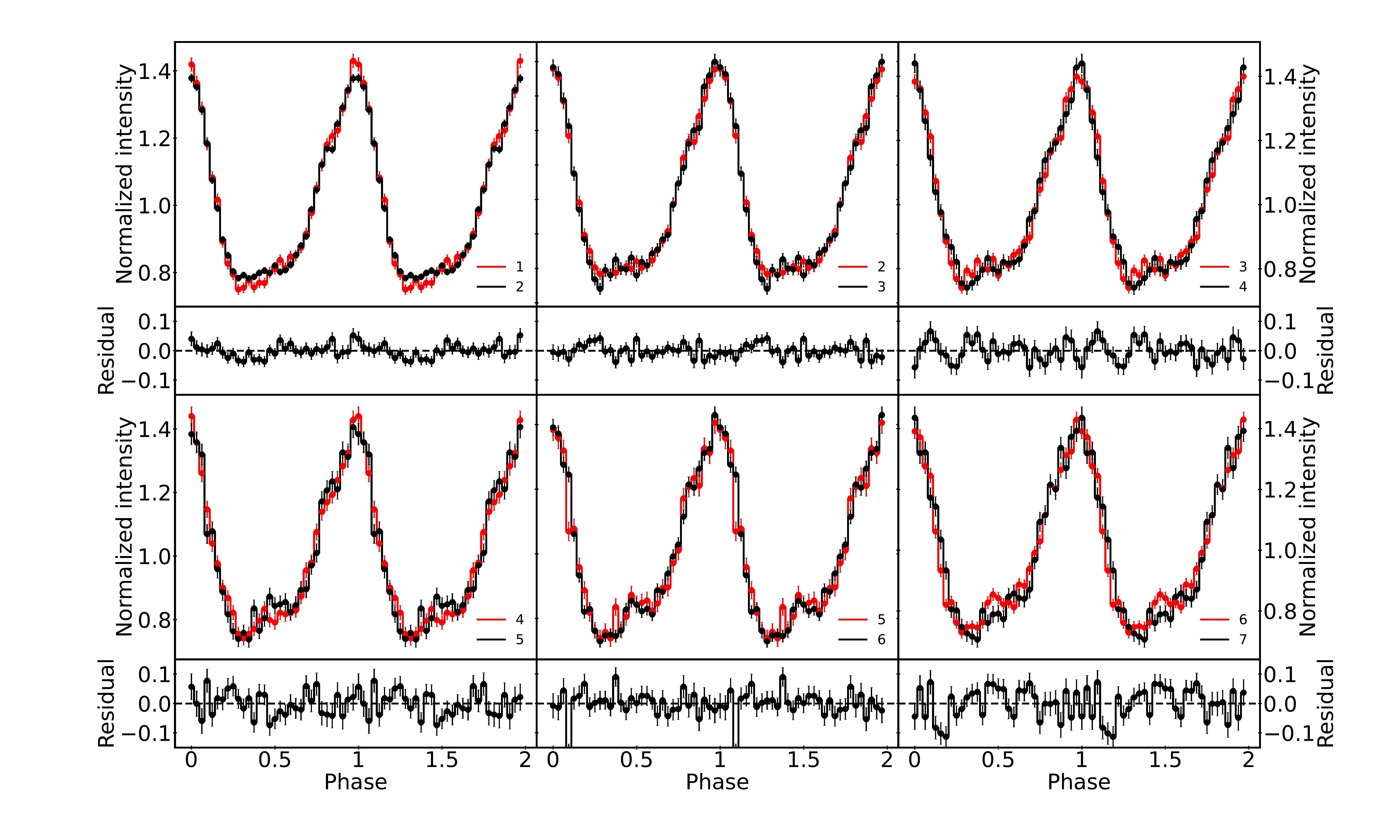}
\caption{Comparison results of normalized pulse profiles of \targetcc . The numbers 1-7 represent the average pulse profiles for the seven time segments, divided based on the vertical lines shown in Figure \ref{timing:c}. The residuals for each pair of normalized pulse profiles, obtained by subtracting one profile from the other, are plotted below the corresponding panels.}
\label{pro:c}
\end{figure*}

A summary of the timing behavior of \targetcc as observed by \nicer is shown in Figure \ref{timing:c}. The long-term timing parameters are listed in Table \ref{table:c}, and the pulse profiles are presented in Figure \ref{pro:c}.

Panels (a) and (b) of figure \ref{timing:c} show that \targetcc underwent multiple timing discontinuities. Four glitches are marked by vertical dashed lines, and a state transition region is highlighted between the vertical solid lines.

\targetcc remains one of the most prolific sources in terms of timing anomalies. During the \nicer observation period, the source exhibits four distinct glitch events. The first event displays the largest frequency change of 137 nHz, while subsequent events show smaller magnitudes. It is noteworthy that a special evolutionary behavior occurs around 59800 MJD, between the two vertical solid lines. At first glance, this behavior resembles two consecutive spin-down glitches. However, upon closer inspection, we identify it as a gradual decrease in spin frequency accompanied by a gradual decline in the frequency derivative -- reminiscent of the slow glitch phenomenon reported by \citet{zhou_2022a}. However, we do not refer to it as a slow anti-glitch, as overly detailed naming conventions may hinder the search for patterns. Instead, we describe this episode as a state transition (encompassing a slow glitch-like event), possibly caused by a change in the torque acting on the star.
A plausible candidate for this state transition is that, assuming the magnetar undergoes frequent jumps caused by superfluid vortex discharges induced by crustquakes, the heating resulting from the dissipation of mechanical and magnetic field energy changes the direction of movement of the vortex lines. The observed sinusoidal-like oscillations in deceleration rate and spin frequency are consistent with the effect of time-varying internal torque applied to the crust, arising from changed configuration of vortex lines \citep{Erbil_2023}.

We divide the dataset into seven segments based on the glitch epochs, including the interval marked by the solid lines, and construct an average pulse profile for each. Each profile consists of 32 phase bins based on the corresponding ephemeris shown in Table \ref{table:c-s}. After normalization, the pulse profiles before and after each event are compared, as shown in Figure \ref{pro:c}. In the 0.5–8 keV energy range, the pulse profile of \targetcc consistently exhibits a stable single-peaked morphology. 
We do not find significant changes in the pulse profiles, except for the last set (which includes only six observations). The results of the chi-squared test are presented in Table \ref{table:P_value}.

\centerwidetable
\tabletypesize{\footnotesize}
\begin{deluxetable*}{lccccccc}
\tablewidth{0pt}
\label{table:c-s}
\tablecaption{Spin parameters of \targetc without glitches}
\startdata
  &  &  \\
Parameters        & R.A.                & Decl.          \\
Value    & 17:08:46.87     & -40:08:52.44  \\
\hline
Parameters  & S1  & S2  & S3  & S4  & S5  & S6  & S7  \\
\hline
MJD range  & 58550-58755  & 58755-59290  & 59290-59535  & 59535-59760    & 59760-59870   & 59870-60200    & 60200-60510  \\
Epoch (MJD)                                             & 58650      & 59000      & 59400      & 59700      & 59800        & 60100        & 60400  \\
$\phi_0$   & 0.063(3)      & 0.734(2)      & 0.957(3)      & 0.398(5)      & 0.676(5)        & 0.483(3)        & 0.222(5)  \\
$\nu $(${\rm Hz}$)                                      & 0.090813358(1)  & 0.0908084107(3)  & 0.0908026601(4)  & 0.090798317(2)  & 0.090796633(3)  & 0.0907921232(5)    & 0.0907878128(5)    \\
$\dot\nu$(${\rm 10^{-13} ~Hz ~s^{-1}}$)               & -1.666(3)    & -1.6881(2)    & -1.695(1)    & -1.679(6)    & -2.27(3)     & -1.6621(9)    & -1.676(1)    \\
$\ddot\nu$(${\rm 10^{-22} ~Hz ~s^{-2}}$)              & 3(1)    & -0.19(4)    & -    & -    & 102(15)    & -    & -    \\
Residuals (ms)                                          & 82.463      & 78.647     & 107.247    & 131.581      & 53.455    & 88.244      & 34.840      \\
\enddata
\tablecomments{S1 to S7 represent the seven time ranges obtained by using the vertical lines in Figure \ref{timing:c} as segmentation points.}
\end{deluxetable*}

\subsection{\targetd} \label{sec:res-4}

\begin{figure*}
\begin{center}
\includegraphics[width=0.65\textwidth,clip]{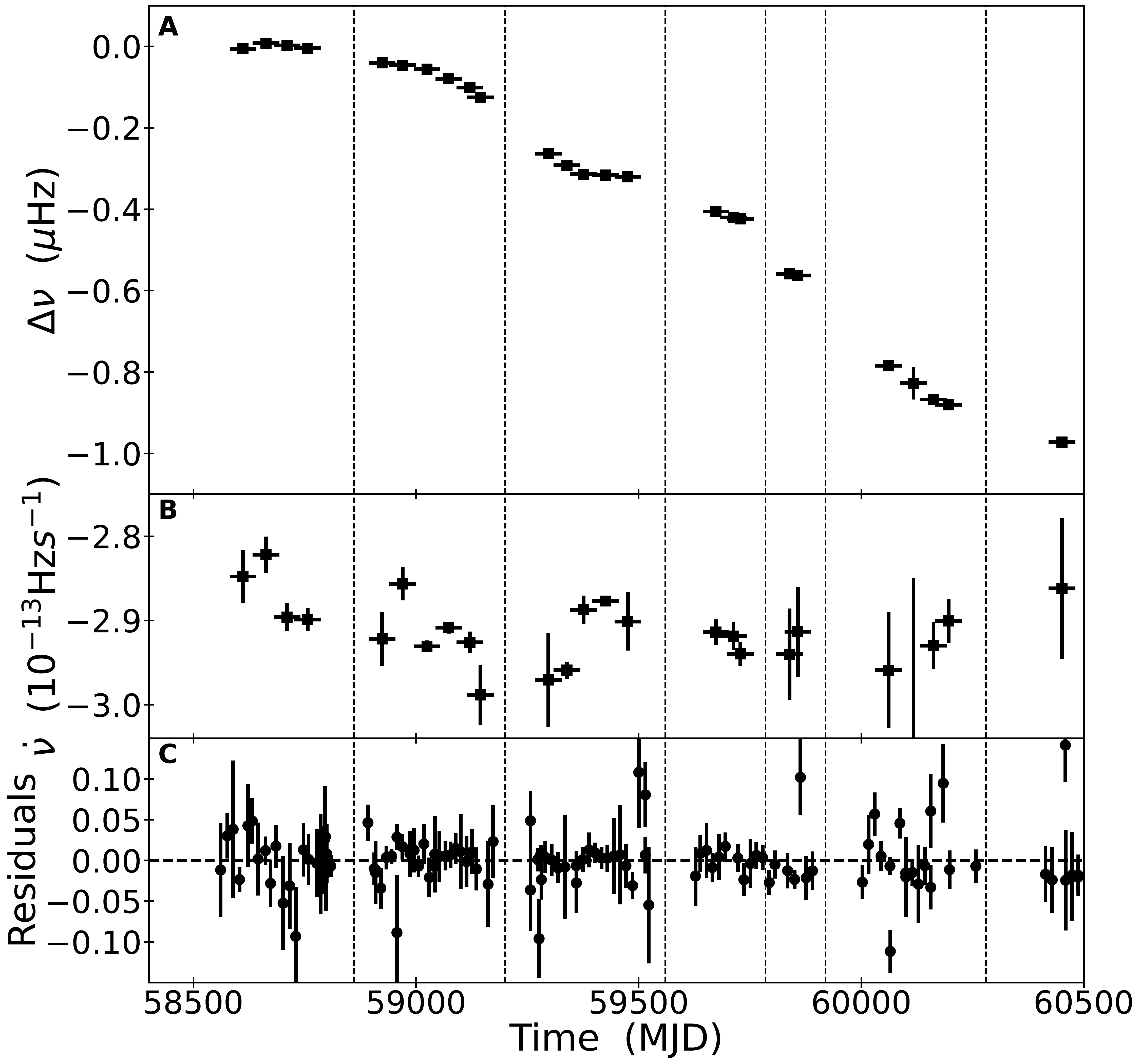}
\caption{The long-term timing evolution of \targetd. Panel(A), Frequency as a function of time with a linear trend in frequency subtracted. Panel(B), Frequency derivative as a function of time. Panel(C), The timing residuals of all TOAs.} 
\label{timing:d}
\end{center}
\end{figure*}

\begin{figure*}
\begin{center}
\includegraphics[width=0.9\textwidth,clip]{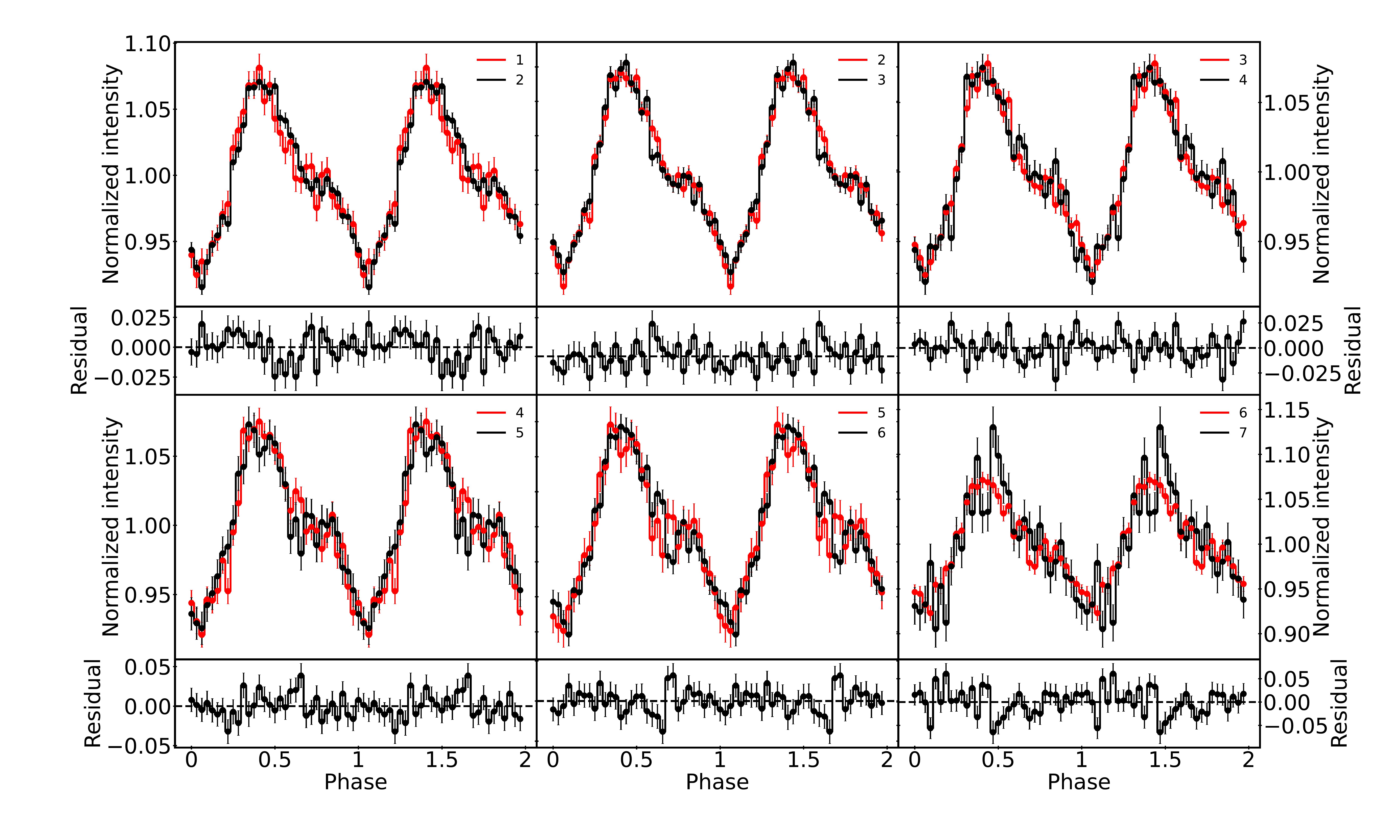}
\caption{Comparison results of normalized pulse profiles of \targetd . The numbers 1-7 represent the average pulse profiles for the seven time segments, divided based on the vertical dashed lines shown in Figure \ref{timing:d}. The residuals for each pair of normalized pulse profiles, obtained by subtracting one profile from the other, are plotted below the corresponding panels.}
\label{pro:d}
\end{center}
\end{figure*}

A summary of the behavior of \targetd as observed by \nicer is shown in Figure \ref{timing:d}. The long-term timing parameters of the source are listed in Table \ref{table:d} and the pulse profiles are presented in Figure \ref{pro:d}.

Panels (a) and (b) of figure \ref{timing:d} reveal that \targetd is a timing-noisy source: its rotational frequency derivative exhibits significant variations on timescales of years.

Due to observational constraints such as Sun-avoidance angles, \nicer's coverage of \targetd is subject to periodic gaps, lasting approximately one-quarter to one-third of each year. Combined with the intrinsic timing noise of the source, this makes timing analysis more challenging than for the other three magnetars in our sample. We do not identify any definitive glitch events during the observation period. Although some apparent discontinuities are present before and after the data gaps, we are unable to determine their precise nature—whether they correspond to actual glitches, gradual state transitions (as in the case of \targetcc), or are simply artifacts of timing noise. Therefore, we divide the data based on these gaps and provide independent ephemerides and corresponding pulse profiles for each segment, as shown in Table \ref{table:d} and Figure \ref{pro:d}. 

It is worth noting that although the fourth vertical dashed line in Figure \ref{timing:d} does not fall within a data gap, it is followed by only seven data points before the next gap, and the folded profile in this region has a low signal-to-noise ratio. Thus, we do not classify this feature as a glitch. Instead, we treat the adjacent segments as independent ephemerides for fitting purposes. Despite the lack of clear glitch identification, the observed long-term trends in spin frequency and its derivative suggest that the rotational behavior of \targetd has undergone irregular changes.

We segment the dataset into seven intervals based on the timing gaps (including the segment associated with the fourth vertical dashed line) and construct an average pulse profile for each.  Each profile consists of 32 bins based on the ephemeris. We normalize and compare the pulse profiles before and after each line, as shown in Figure \ref{pro:d}. The pulse profile of \targetd is characterized by a relatively stable single-peaked structure, accompanied by a small bump on the trailing edge.  
We do not find significant changes in the pulse profiles and the results of the chi-squared test are presented in Table \ref{table:P_value}.

\begin{figure}
\begin{center}
\includegraphics[width=0.4\textwidth,clip]{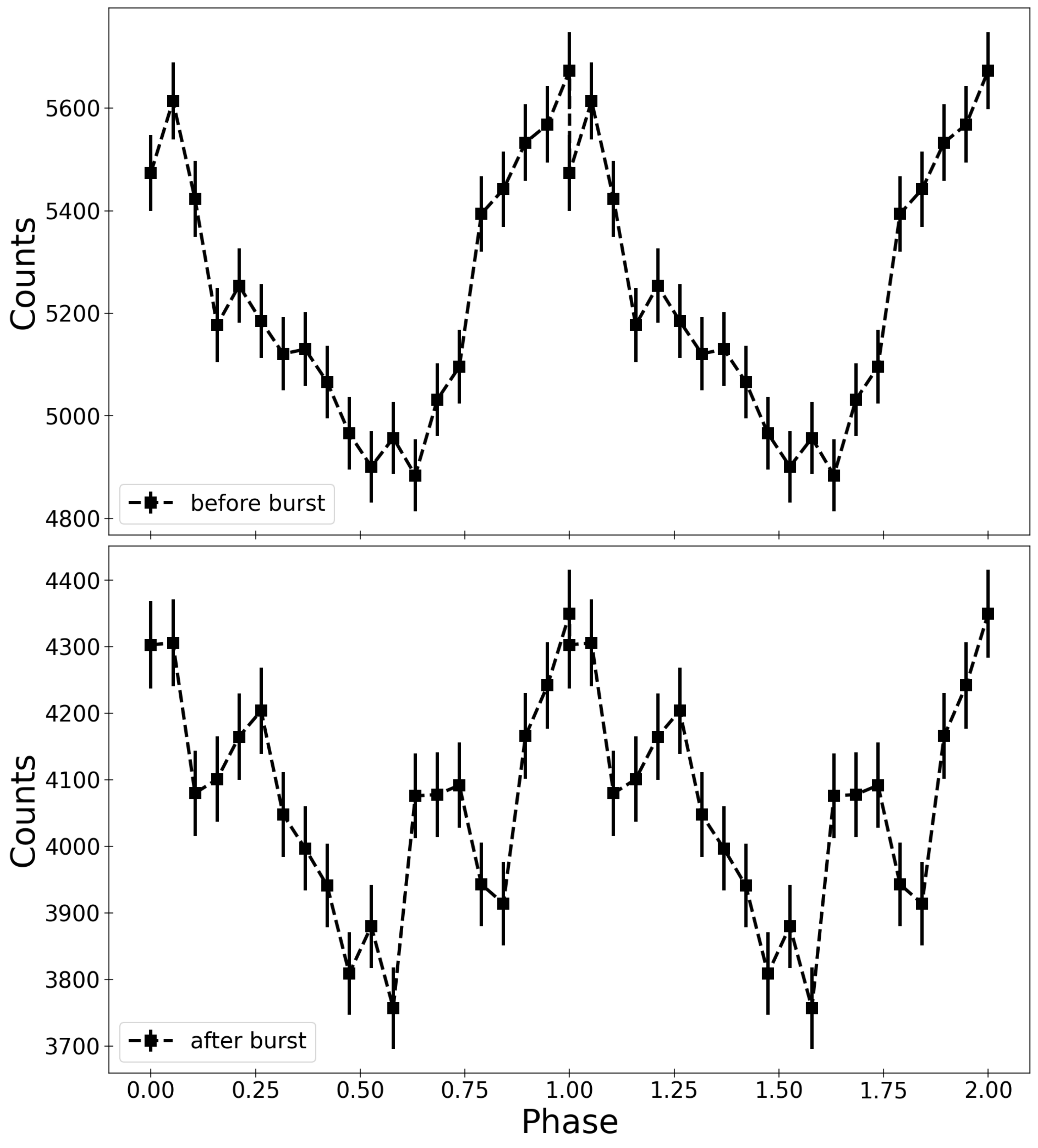}
\caption{The pulse profile of \targetd before (top) and after (bottom) burst, with 20 phase bins. The burst occurred on August 21, 2024, at MJD 60543 \citep{Dichiara_2024}.}
\label{pro:d-burst}
\end{center}
\end{figure}

Importantly, following the time span analyzed here, \targetd entered an active phase, during which an outburst occurred, accompanied by a short burst, a flux enhancement, and a glitch. Notably, its pulse profile  underwent a dramatic transformation. Unlike the gradual evolutionary changes observed earlier, the profile shifted from a quasi-single-peaked to a complex multi-peaked morphology, as illustrated in Figure \ref{pro:d-burst}. For detailed timing and spectral analysis of this event, refer to the report by \citet{Younes_2025}.  Additional analysis by Fu et al. (2026, in preparation) will be presented in a forthcoming publication.

\centerwidetable
\tabletypesize{\footnotesize}
\begin{deluxetable*}{lccccccc}
\tablewidth{0pt}
\label{table:d}
\tablecaption{Spin parameters of \targetd}
\startdata
  &  &  \\
Parameters        & R.A.                & Decl.          \\
Value    & 18:41:19.343     & -04:56:11.16  \\
\hline
Parameters  & S1  & S2  & S3  & S4  & S5  & S6  & S7  \\
\hline
MJD range  & 58530-58860  & 58860-59200  & 59230-59570  & 59600-59790    & 59790-59910   & 59980-60290    & 60410-60490  \\
Epoch (MJD)                                             & 58623      & 59000      & 59400      & 59700      & 59850        & 60050        & 60450  \\
$\phi_0$   & 0.026(9)      & 0.727(5)      & 0.058(7)      & 0.625(7)      & 0.066(9)        & 0.539(7)        & 0.84(1)  \\
$\nu $(${\rm Hz}$)                                      & 0.084748357(2)  & 0.084738944(1)  & 0.084728741(1)  & 0.084721187(1)  & 0.084717311(2)  & 0.084712131(3)    & 0.0847019978(5)    \\
$\dot\nu$(${\rm 10^{-13} ~Hz ~s^{-1}}$)               & -2.85(1)    & -2.903(2)    & -2.900(1)    & -2.921(5)    & -2.94(1)     & -2.99(1)    & -2.936(6)    \\
$\ddot\nu$(${\rm 10^{-22} ~Hz ~s^{-2}}$)           & 3(1)    & -5.0(7)    & 6(1)    & -    & -    & 5(1)    & -    \\
Residuals (ms)                                          & 280.999      & 183.306     & 222.007    & 134.321      & 92.078    & 299.326      & 22.906      \\
\enddata
\tablecomments{S1 to S7 represent the seven time ranges obtained by using the vertical lines in Figure \ref{timing:d} as segmentation points.}
\end{deluxetable*}

\begin{deluxetable}{lcccccc}
\tablewidth{0pt}
\label{table:P_value}
\tablecaption{The P-value of the chi-square test for pulse profile residuals}
\startdata
  &  &  \\
\targeta           & S1-S2                & S2-S3          \\
$\chi^{2}$ test    & 0.0024     & 2.2e-8       \\
% K-S test           & 0.999                & 0.997   \\
\hline
\targetb           & S1-S2                & S2-S3          \\
$\chi^{2}$ test    & 1.2e-6     & 9.3e-8       \\
% K-S test           & 0.997                & 0.966                  \\
\hline
\targetcc          & S1-S2      & S2-S3      & S3-S4      & S4-S5      & S5-S6      & S6-S7       \\
$\chi^{2}$ test    & 0.090     & 0.175      & 0.118          & 0.155          & 0.046      & 0.00010  \\
% K-S test           & 0.635                & 0.968                 & 0.999          & 0.837          & 0.968      & 0.838 \\
\hline
\targetd          & S1-S2      & S2-S3      & S3-S4      & S4-S5      & S5-S6      & S6-S7       \\
$\chi^{2}$ test    & 0.294     & 0.578      & 0.073          & 0.506          & 0.455      & 0.020  \\
% K-S test           & 0.999                & 0.999                 & 0.999          & 0.968          & 0.968      & 0.968 \\
\hline
\enddata
\tablecomments{S1-S7 represent the pulse profile in the pulse profile diagram of each source.}
\end{deluxetable}

\section{DISCUSSION AND SUMMARY} \label{sec:dis}

\centerwidetable
\tabletypesize{\footnotesize}
\begin{deluxetable*}{cccccc}
\tablewidth{0pt}
\label{table:statistics}
\tablecaption{Timing Events of Four AXPs Observed by \nicer}
\startdata
  &  &  \\
Source         & Timing Events         & Epoch (MJD)      & $\triangle\nu_{g}$ (Hz)       & Profile Change                 & Reference     \\
\hline
\targeta       & glitch                & 57934.48         & $1.78(2) \times 10^{-7}$      & -                              & \citep{Younes_2020}  \\
\targeta       & anti-glitch           & 58575            & $-8.95(5) \times 10^{-8}$     & Y \tablenotemark{b}                              & This work \tablenotemark{a}, \citep{Younes_2020}  \\
\targeta       & glitch                & 59780            & $1.935(2) \times 10^{-7}$     & Y                              & This work  \\
\hline
\targetb       & anti-glitch           & 59540            & $-1.25(9) \times 10^{-8}$     &  Y     & This work  \\
\targetb       & anti-glitch           & 59850            & $-5.42(9) \times 10^{-8}$     &  Y                      & This work  \\
\hline
\targetcc       & glitch               & 58755            & $1.37(2) \times 10^{-7}$      & N                               & This work  \\ 
\targetcc       & glitch               & 59290            & $8.4(2) \times 10^{-8}$       & N                               & This work  \\
\targetcc       & glitch               & 59540            & $2.9(6) \times 10^{-8}$       & N                               & This work  \\
\targetcc       & glitch               & 60200            & $2.2(3) \times 10^{-8}$       & Y                               & This work  \\
\targetcc       & state transition     & Around 59800     & -                             & N                               & This work  \\
\hline
\targetd       & glitch                & 60543            & $6.1(4) \times 10^{-8}$       & Y \tablenotemark{c}        & \citep{Younes_2025}  \\
\enddata
% \tablecomments{}
\tablenotetext{a}{Accompanied by an exponential decay term with $\tau_{d} = 89(17)$ Days.}
\tablenotetext{b}{Possible signs of evolution, but not changes in the pulse profile characteristics.}
\tablenotetext{c}{The pulse profile undergoes a significant characteristic change (from a quasi-single-peaked to a complex multi-peaked morphology), accompanied by an outburst.}
\end{deluxetable*}

We present, for the first time, a systematic timing and pulse profile analysis from \nicer's regular monitoring of four AXPs (\targeta, \targetb, \targetcc, and \targetd) over roughly 7 years. This has yielded a rich data set, significantly expanding the sample of glitch events in magnetars.  During this time span, each source exhibited its own unique rotational behavior. For example, \targeta is a very stable rotator, while \targetcc, in contrast, is a frequent glitcher, and \targetb stands out even more due to its anti-glitch behavior. That said, even among magnetars, each source retains its distinct characteristics in terms of radiation and spin evolution. Systematic analysis not only reveals clear common behaviors but also highlights differences between sources, which can help explore the underlying triggering mechanisms. This is crucial for developing a coherent theory to explain magnetar phenomenology.

Specifically, we identify a total of 10 timing events. In addition to one glitch and one anti-glitch in \targeta that have already been reported \citep{Younes_2020}, we discover 8 new events: \targeta exhibits another glitch, and in the previous anti-glitch event, we find an exponential decay term, which was not recognized in earlier reports. \targetb shows two spin-down glitches. \targetcc undergoes four glitches and one unique state transition event. Due to observational gaps, we do not identify specific glitch events in \targetd. Detailed statistics of the timing events are presented in Table \ref{table:statistics}. 
% Note: After the range of data we analyzed, \targetd enters an active phase accompanied by a glitch event, which \citet{Younes_2025} analyzed and reported in detail. Therefore, we include this event in Table \ref{table:statistics} to provide a complete statistics of timing events within \nicer's monitoring range.
\footnote{After the range of data we analyzed, \targetd enters an active phase accompanied by a glitch event, which \citet{Younes_2025} analyzed and reported in detail. Therefore, we include this event in Table \ref{table:statistics} to provide a complete statistics of timing events within \nicer's monitoring range.}

Regarding the timing results, there are three thought-provoking points aside from ordinary glitch behaviors. (1) From the current statistics of timing events, the glitch events significantly outnumber the anti-glitch events, which is consistent with previous results. The two new anti-glitch events both occurred in \targetb, making this source the one with the highest number of such events. Moreover, in this source, anti-glitch events occur more frequently than glitch events, which clearly contradicts the overall statistics of magnetars. What makes this source so unique remains worth further exploration, and exploring the factors behind this difference is crucial for understanding the underlying physical mechanisms. 
Presumably, the strong crustal toroidal magnetic field contained within the source generates stresses that cause the superfluid vortex lines to move inward, therby producing anti-glitches.
(2) A state transition event occurs in the source \targetcc . This event is similar to the ``slow glitch" reported by \citet{zhou_2022a}, but instead of a spin-up, it involves a spin-down. Whether this gradually evolving event is caused by changes in the magnetic field's torque and whether it can further constrain theoretical models of magnetars is also worth exploring. 
Whether this gradually evolving event is caused by changes in the coupling between the external magnetospheric and internal superfluid torques, possibly associated with temperature variations and magnetic-field topology changes that may accompany magnetar bursting activity, and whether it can further constrain theoretical models of magnetars is also worth exploring.
(3) The anti-glitch event in \targeta is accompanied by an exponential recovery term. This is similar to \targetb, PSR J1846-0258, and PSR J1119-6127 but differs from SGR J1900+14 and SGR J1935+2154, which show no exponential term \citep{Ge_2024}. Regardless, most glitch events are not accompanied by an exponential term. What causes this behavior and how the exponential component may provide for theoretical models are questions that also merit further exploration.  
One possible explanation for this exponential relaxation after the anti-glitch event is the response of the vortex lines, which creep against the toroidal-arranged flux tubes in the magnetar's core region, to changes in the crust's rotation rate.

Also, we study the evolution of the pulse profiles and find that the profiles of \targeta and \targetb both evolve. This is consistent with the earlier conclusion that evolution appears to be a generic property of magnetars \citep{Dib_2008, Dib_2014}. However, this type of evolution occurs at a low level and requires the accumulation of sufficiently long pulse profiles to become evident. For \targetcc and \targetd, the lack of significant changes in the pulse profiles, compared to those of \targeta and \targetb, may be due to an insufficient signal-to-noise ratio. We calculated the chi-squared values for the significance of the profiles for the latter two, which can be more than ten times smaller than those for the former.

In 2020, a FRB was discovered to be associated with SGR 1935+2154 \citep{Andersen_2020, Bochenek_2020}, leading to the widespread acceptance that magnetars are one of the central engine for the origin for FRBs..  \citet{Younes_2023} observed that following a spin-down glitch, SGR 1935+2154 emitted three FRB-like radio bursts, followed by pulsed radio radiation lasting for a month. The rarity and near-synchronization of these events suggest a possible connection, providing critical clues about their origin and triggering mechanisms. Additionally, in October 2022, another FRB event occurred from this source, accompanied by two glitch events within nine hours before and after the FRB \citep{Hu_2024}. The first glitch was associated with a significant increase in burst rate and X-ray flux. However, how magnetars produce FRBs and whether there is a relationship between glitches and FRBs remain unclear. It is worth mentioning that magnetars with radio emissions are very rare, and none of the four magnetars discussed in this paper exhibit radio emission \citep{Burgay_2007, Lu_2024, Bai_2025}.

With the improvement of observational capabilities, we have uncovered many phenomenological events related to magnetars. However, these events also raise numerous questions about the physical mechanisms operating in magnetars and even in rotation-powered pulsars. The discovery of anti-glitches in rotation-powered pulsars indicates that such phenomena are not exclusive to magnetars \citep{Zhou_2024, Tuo_2024}. Additionally, two rotation-powered pulsars have now exhibited magnetar-like behaviors, such as activity accompanied by glitches \citep{Dai_2018, Archibald_2018, Sathyaprakash_2024, Hu_2023}. These observations lead us to consider whether a unified theory of pulsar glitches exists and what role the magnetic field plays in the activity of pulsars. Finally, the current sample of magnetar timing events is still insufficient, and whether there are more peculiar timing behaviors and whether our current statistical understanding holds as the sample size increases remains unknown. Continued systematic monitoring to obtain more samples, especially of rare events, is key to addressing these issues. Excitingly, China is set to launch the enhanced X-ray Timing and Polarimetry mission(eXTP) in the near future \citep{Zhang_2025}, which will simultaneously provide imaging, timing, and polarimetric analysis of sources \citep{Ge_2025}. Its unique capabilities are expected to make significant contributions to understanding the physics of magnetars.

\begin{acknowledgments}
The authors thank supports from the National Natural Science Foundation of China under Grants 12473041,12373051. This work is partially supported by the China Manned Space Program with grant no CMS-CSST-2025-A13, the National Key R\&D Program of China (2021YFA0718500) from the Minister of Science and Technology of China (MOST) and the China's Space Origins Exploration Program.  
S.Q.Z. acknowledges support from the Key Project of the Sichuan Science and Technology Education Joint Fund (25LHJJ0097) and the PhD Start-up Fund of China West Normal University (25KE033). EG is supported by the Doctor Foundation of Qingdao Binhai University (No. BJZA2025025).
\end{acknowledgments}

% \clearpage
%\bibliography{main}{}

\begin{thebibliography}{}
\expandafter\ifx\csname natexlab\endcsname\relax\def\natexlab#1{#1}\fi
\providecommand{\url}[1]{\href{#1}{#1}}
\providecommand{\dodoi}[1]{doi:~\href{http://doi.org/#1}{\nolinkurl{#1}}}
\providecommand{\doeprint}[1]{\href{http://ascl.net/#1}{\nolinkurl{http://ascl.net/#1}}}
\providecommand{\doarXiv}[1]{\href{https://arxiv.org/abs/#1}{\nolinkurl{https://arxiv.org/abs/#1}}}

% type= article
\bibitem[{O. {Akbal} {et~al.}(2015){Akbal}, {G{\"u}gercino{\u{g}}lu},
  {{\c{S}}a{\c{s}}maz Mu{\c{s}}}, \& {Alpar}}]{Akbal_2015}
{Akbal}, O., {G{\"u}gercino{\u{g}}lu}, E., {{\c{S}}a{\c{s}}maz Mu{\c{s}}}, S.,
  \& {Alpar}, M.~A. 2015, \bibinfo{title}{{Peculiar glitch of PSR J1119-6127
  and extension of the vortex creep model},} \mnras, 449, 933,
  \dodoi{10.1093/mnras/stv322}

% type= article
\bibitem[{H. An {et~al.}(2015)An, Archibald, Hascoët, Kaspi, Beloborodov,
  Archibald, Beardmore, Boggs, Christensen, Craig, Gehrels, Hailey, Harrison,
  Kennea, Kouveliotou, Stern, Younes, \& Zhang}]{An_2015}
An, H., Archibald, R.~F., Hascoët, R., {et~al.} 2015, \bibinfo{title}{DEEP
  NuSTAR AND SWIFT MONITORING OBSERVATIONS OF THE MAGNETAR 1E 1841-045,} The
  Astrophysical Journal, 807, 93, \dodoi{10.1088/0004-637X/807/1/93}

% type= article
\bibitem[{B. Andersen {et~al.}(2020)Andersen, Bandura, \&
  Bhardwaj}]{Andersen_2020}
Andersen, B., Bandura, K., \& Bhardwaj, M. 2020, \bibinfo{title}{A bright
  millisecond-duration radio burst from a Galactic magnetar,} Nature, 587, 54,
  \dodoi{10.1038/s41586-020-2863-y}

% type= article
\bibitem[{P.~W. {Anderson} \& N. {Itoh}(1975){Anderson} \&
  {Itoh}}]{Anderson_1975}
{Anderson}, P.~W., \& {Itoh}, N. 1975, \bibinfo{title}{Pulsar glitches and
  restlessness as a hard superfluidity phenomenon,} Nature, 256, 25,
  \dodoi{10.1038/256025a0}

% type= article
\bibitem[{D. Antonopoulou {et~al.}(2022)Antonopoulou, Haskell, \&
  Espinoza}]{Antonopoulou_2022}
Antonopoulou, D., Haskell, B., \& Espinoza, C.~M. 2022, \bibinfo{title}{Pulsar
  glitches: observations and physical interpretation,} Reports on Progress in
  Physics, 85, 126901, \dodoi{10.1088/1361-6633/ac9ced}

% type= article
\bibitem[{R.~F. Archibald {et~al.}(2017)Archibald, Kaspi, Scholz, Beardmore,
  Gehrels, \& Kennea}]{Archibald_2017}
Archibald, R.~F., Kaspi, V.~M., Scholz, P., {et~al.} 2017,
  \bibinfo{title}{SWIFT OBSERVATIONS OF TWO OUTBURSTS FROM THE MAGNETAR 4U
  0142+61,} The Astrophysical Journal, 834, 163,
  \dodoi{10.3847/1538-4357/834/2/163}

% type= article
\bibitem[{R.~F. Archibald {et~al.}(2018)Archibald, Kaspi, Tendulkar, \&
  Scholz}]{Archibald_2018}
Archibald, R.~F., Kaspi, V.~M., Tendulkar, S.~P., \& Scholz, P. 2018,
  \bibinfo{title}{The 2016 Outburst of PSR J1119-6127: Cooling and a
  Spin-down-dominated Glitch,} The Astrophysical Journal, 869, 180,
  \dodoi{10.3847/1538-4357/aaee73}

% type= article
\bibitem[{R.~F. {Archibald} {et~al.}(2013){Archibald}, {Kaspi}, {Ng},
  {Gourgouliatos}, {Tsang}, {Scholz}, {Beardmore}, {Gehrels}, \&
  {Kennea}}]{Archibald_2013}
{Archibald}, R.~F., {Kaspi}, V.~M., {Ng}, C.~Y., {et~al.} 2013,
  \bibinfo{title}{{An anti-glitch in a magnetar},} \nat, 497, 591,
  \dodoi{10.1038/nature12159}

% type= article
\bibitem[{J. Bai {et~al.}(2025)Bai, Wang, Dai, Wang, Yuan, Yan, Shang, Xu,
  Dang, \& Zhang}]{Bai_2025}
Bai, J., Wang, N., Dai, S., {et~al.} 2025, \bibinfo{title}{Deep Searches for
  Radio Pulsations and Bursts from Four Magnetar and a Magnetar-like Pulsar
  with FAST,} The Astrophysical Journal, 979, 122,
  \dodoi{10.3847/1538-4357/ada3c4}

% type= article
\bibitem[{L. Baldini {et~al.}(2022)Baldini, Bucciantini, Lalla, Ehlert,
  Manfreda, Negro, Omodei, Pesce-Rollins, SgrÃ², \& Silvestri}]{Baldini_2022}
Baldini, L., Bucciantini, N., Lalla, N.~D., {et~al.} 2022,
  \bibinfo{title}{ixpeobssim: A simulation and analysis framework for the
  imaging X-ray polarimetry explorer,} SoftwareX, 19, 101194,
  \dodoi{https://doi.org/10.1016/j.softx.2022.101194}

% type= article
\bibitem[{C.~D. Bochenek {et~al.}(2020)Bochenek, Ravi, \&
  Belov}]{Bochenek_2020}
Bochenek, C.~D., Ravi, V., \& Belov, K.~V. 2020, \bibinfo{title}{A fast radio
  burst associated with a Galactic magnetar,} Nature, 587, 59,
  \dodoi{10.1038/s41586-020-2872-x}

% type= article
\bibitem[{M. Burgay {et~al.}(2007)Burgay, Rea, \& Israel}]{Burgay_2007}
Burgay, M., Rea, N., \& Israel, G. e.~a. 2007, \bibinfo{title}{Search for radio
  pulsations in four anomalous X-ray pulsars and discovery of two new pulsars,}
  Astrophysics and Space Science, 308, 521.
\newblock \url{https://doi.org/10.1007/s10509-007-9353-7}

% type= article
\bibitem[{F. Coti~Zelati {et~al.}(2017)Coti~Zelati, Rea, Pons, Campana, \&
  Esposito}]{Francesco_2017}
Coti~Zelati, F., Rea, N., Pons, J.~A., Campana, S., \& Esposito, P. 2017,
  \bibinfo{title}{Systematic study of magnetar outbursts,} Monthly Notices of
  the Royal Astronomical Society, 474, 961, \dodoi{10.1093/mnras/stx2679}

% type= article
\bibitem[{S. Dai {et~al.}(2018)Dai, Johnston, Weltevrede, Kerr, Burgay,
  Esposito, Israel, Possenti, Rea, \& Sarkissian}]{Dai_2018}
Dai, S., Johnston, S., Weltevrede, P., {et~al.} 2018, \bibinfo{title}{Peculiar
  spin frequency and radio profile evolution of PSR J1119â6127 following
  magnetar-like X-ray bursts,} Monthly Notices of the Royal Astronomical
  Society, 480, 3584, \dodoi{10.1093/mnras/sty2063}

% type= article
\bibitem[{R. {Dib} \& V.~M. {Kaspi}(2014){Dib} \& {Kaspi}}]{Dib_2014}
{Dib}, R., \& {Kaspi}, V.~M. 2014, \bibinfo{title}{{16 yr of RXTE Monitoring of
  Five Anomalous X-Ray Pulsars},} \apj, 784, 37,
  \dodoi{10.1088/0004-637X/784/1/37}

% type= article
\bibitem[{R. Dib {et~al.}(2008)Dib, Kaspi, \& Gavriil}]{Dib_2008}
Dib, R., Kaspi, V.~M., \& Gavriil, F.~P. 2008, \bibinfo{title}{Glitches in
  Anomalous X-Ray Pulsars,} The Astrophysical Journal, 673, 1044,
  \dodoi{10.1086/524653}

% type= article
\bibitem[{S. Dichiara \& D.~M. Palmer(2024)Dichiara \& Palmer}]{Dichiara_2024}
Dichiara, S., \& Palmer, D.~M. 2024, \bibinfo{title}{Re-activation of Soft
  Gamma Repeater SGR 1E 1841-045 / Kes 73,} The Astronomer's Telegram, 16784, 1

% type= inbook
\bibitem[{P. Esposito {et~al.}(2021)Esposito, Rea, \& Israel}]{Esposito_2021}
Esposito, P., Rea, N., \& Israel, G.~L. 2021, Magnetars: A Short Review and
  Some Sparse Considerations, ed. T.~M. Belloni, M.~M{\'e}ndez, \& C.~Zhang
  (Berlin, Heidelberg: Springer Berlin Heidelberg), 97--142,
  \dodoi{10.1007/978-3-662-62110-3_3}

% type= article
\bibitem[{G.~G. {Fahlman} \& P.~C. {Gregory}(1981){Fahlman} \&
  {Gregory}}]{Fahlman_1981}
{Fahlman}, G.~G., \& {Gregory}, P.~C. 1981, \bibinfo{title}{{An X-ray pulsar in
  SNR G109.1-1.0},} \nat, 293, 202, \dodoi{10.1038/293202a0}

% type= article
\bibitem[{R.~D. Ferdman {et~al.}(2015)Ferdman, Archibald, \&
  Kaspi}]{Ferdman_2015}
Ferdman, R.~D., Archibald, R.~F., \& Kaspi, V.~M. 2015,
  \bibinfo{title}{LONG-TERM TIMING AND EMISSION BEHAVIOR OF THE YOUNG CRAB-LIKE
  PULSAR PSR B0540â69,} The Astrophysical Journal, 812, 95,
  \dodoi{10.1088/0004-637X/812/2/95}

% type= article
\bibitem[{F.~P. Gavriil {et~al.}(2011a)Gavriil, Dib, \& Kaspi}]{Gavriil_2011a}
Gavriil, F.~P., Dib, R., \& Kaspi, V.~M. 2011a, \bibinfo{title}{THE 2006â2007
  ACTIVE PHASE OF ANOMALOUS X-RAY PULSAR 4U 0142+61: RADIATIVE AND TIMING
  CHANGES, BURSTS, AND BURST SPECTRAL FEATURES,} The Astrophysical Journal,
  736, 138, \dodoi{10.1088/0004-637X/736/2/138}

% type= article
\bibitem[{F.~P. {Gavriil} {et~al.}(2002){Gavriil}, {Kaspi}, \&
  {Woods}}]{Gavriil2002_na}
{Gavriil}, F.~P., {Kaspi}, V.~M., \& {Woods}, P.~M. 2002,
  \bibinfo{title}{Magnetar-like X-ray bursts from an anomalous X-ray pulsar,}
  Nature, 419, 142, \dodoi{10.1038/nature01011}

% type= article
\bibitem[{M. Ge {et~al.}(2022)Ge, Yang, Lu, Zhou, Ji, Zhang, Zhang, Zhang,
  Wang, Lee, Zhu, Li, Hou, \& Li}]{GE_2022}
Ge, M., Yang, Y.-P., Lu, F., {et~al.} 2022, \bibinfo{title}{A giant glitch from
  the magnetar SGR J1935+2154 before FRB 200428,} \doarXiv{2211.03246}

% type= article
\bibitem[{M. {Ge} {et~al.}(2025){Ge}, {Ji}, {Taverna}, {Tsygankov}, {Xu},
  {Santangelo}, {Zane}, {Zhang}, {Feng}, {Chen}, {Cheng}, {Hou}, {Imbrogno},
  {Israel}, {Kelly}, {Kong}, {Liu}, {Mushtukov}, {Poutanen}, {Suleimanov},
  {Tao}, {Tong}, {Turolla}, {Wang}, {Ye}, {Zhao}, {Geng}, {Lin}, {Wang}, {Xie},
  {Xiong}, {Zhang}, {Fu}, {Lai}, {Li}, {Li}, {Li}, {Li}, {Liu}, {Liu}, {Peng},
  {Shui}, {Tuo}, {Wang}, {Wang}, {Weng}, {You}, {Zheng}, \& {Zhou}}]{Ge_2025}
{Ge}, M., {Ji}, L., {Taverna}, R., {et~al.} 2025, \bibinfo{title}{{Physics of
  Strong Magnetism with eXTP},} arXiv e-prints, arXiv:2506.08369,
  \dodoi{10.48550/arXiv.2506.08369}

% type= article
\bibitem[{M.~Y. Ge {et~al.}(2012)Ge, Lu, Qu, Zheng, Chen, \& Han}]{Ge_2012}
Ge, M.~Y., Lu, F.~J., Qu, J.~L., {et~al.} 2012, \bibinfo{title}{X-RAY
  PHASE-RESOLVED SPECTROSCOPY OF PSRs B0531+21, B1509â58, AND B0540â69 WITH
  RXTE,} The Astrophysical Journal Supplement Series, 199, 32,
  \dodoi{10.1088/0067-0049/199/2/32}

% type= article
\bibitem[{M.-Y. Ge {et~al.}(2024)Ge, Yang, Lu, Zhou, Ji, Zhang, Zhang, Zhang,
  Wang, Lee, Zhu, Li, Hou, \& Li}]{Ge_2024}
Ge, M.-Y., Yang, Y.-P., Lu, F.-J., {et~al.} 2024, \bibinfo{title}{Spin
  Evolution of the Magnetar SGR J1935+2154,} Research in Astronomy and
  Astrophysics, 24, 015016, \dodoi{10.1088/1674-4527/ad0f0c}

% type= inproceedings
\bibitem[{K.~C. Gendreau {et~al.}(2016)Gendreau, Arzoumanian, \&
  Adkins}]{Gendreau_2016}
Gendreau, K.~C., Arzoumanian, Z., \& Adkins, P.~W. 2016, \bibinfo{title}{{The
  Neutron star Interior Composition Explorer (NICER): design and development},}
  in Space Telescopes and Instrumentation 2016: Ultraviolet to Gamma Ray, ed.
  J.-W.~A. den Herder, T.~Takahashi, \& M.~Bautz, Vol. 9905, International
  Society for Optics and Photonics (SPIE), 99051H, \dodoi{10.1117/12.2231304}

% type= article
\bibitem[{R. {Giacconi} {et~al.}(1972){Giacconi}, {Murray}, {Gursky},
  {Kellogg}, {Schreier}, \& {Tananbaum}}]{Giacconi_1972}
{Giacconi}, R., {Murray}, S., {Gursky}, H., {et~al.} 1972, \bibinfo{title}{The
  Uhuru catalog of X-ray sources,} \apj, 182, 281, \dodoi{10.1086/151790}

% type= article
\bibitem[{E.
  {G{\"u}gercino{\u{g}}lu}(2017){G{\"u}gercino{\u{g}}lu}}]{Erbil_2017a}
{G{\"u}gercino{\u{g}}lu}, E. 2017, \bibinfo{title}{{Post-glitch exponential
  relaxation of radio pulsars and magnetars in terms of vortex creep across
  flux tubes},} \mnras, 469, 2313, \dodoi{10.1093/mnras/stx985}

% type= article
\bibitem[{E. {G{\"u}gercino{\u{g}}lu} \& M.~A.
  {Alpar}(2016){G{\"u}gercino{\u{g}}lu} \& {Alpar}}]{Erbil_2016}
{G{\"u}gercino{\u{g}}lu}, E., \& {Alpar}, M.~A. 2016,
  \bibinfo{title}{{Microscopic vortex velocity in the inner crust and outer
  core of neutron stars},} \mnras, 462, 1453, \dodoi{10.1093/mnras/stw1758}

% type= article
\bibitem[{E. {G{\"u}gercino{\u{g}}lu} {et~al.}(2023){G{\"u}gercino{\u{g}}lu},
  {K{\"o}ksal}, \& {G{\"u}ver}}]{Erbil_2023}
{G{\"u}gercino{\u{g}}lu}, E., {K{\"o}ksal}, E., \& {G{\"u}ver}, T. 2023,
  \bibinfo{title}{{On the peculiar rotational evolution of PSR B0950+08},}
  \mnras, 518, 5734, \dodoi{10.1093/mnras/stac3516}

% type= article
\bibitem[{E. {G{\"u}gercino{\v{g}}lu} \& M.~A.
  {Alpar}(2017){G{\"u}gercino{\v{g}}lu} \& {Alpar}}]{Erbil_2017b}
{G{\"u}gercino{\v{g}}lu}, E., \& {Alpar}, M.~A. 2017, \bibinfo{title}{{Neutron
  star dynamics under time-dependent external torques},} \mnras, 471, 4827,
  \dodoi{10.1093/mnras/stx1937}

% type= article
\bibitem[{E. {G{\"u}gercino{\v{g}}lu} \& M.~A.
  {Alpar}(2019){G{\"u}gercino{\v{g}}lu} \& {Alpar}}]{Erbil_2019}
{G{\"u}gercino{\v{g}}lu}, E., \& {Alpar}, M.~A. 2019, \bibinfo{title}{{The
  largest Crab glitch and the vortex creep model},} \mnras, 488, 2275,
  \dodoi{10.1093/mnras/stz1831}

% type= article
\bibitem[{G.~B. {Hobbs} {et~al.}(2006){Hobbs}, {Edwards}, \&
  {Manchester}}]{Hobbs_2006}
{Hobbs}, G.~B., {Edwards}, R.~T., \& {Manchester}, R.~N. 2006,
  \bibinfo{title}{{TEMPO2, a new pulsar-timing package - I. An overview},}
  \mnras, 369, 655, \dodoi{10.1111/j.1365-2966.2006.10302.x}

% type= article
\bibitem[{C.-P. Hu {et~al.}(2024)Hu, Narita, Enoto, \& Younes}]{Hu_2024}
Hu, C.-P., Narita, T., Enoto, T., \& Younes, G. 2024, \bibinfo{title}{Rapid
  spin changes around a magnetar fast radio burst,} Nature, 626, 500,
  \dodoi{10.1038/s41586-023-07012-5}

% type= article
\bibitem[{C.-P. Hu {et~al.}(2023)Hu, Kuiper, Harding, Younes, Blumer, Ho,
  Enoto, Espinoza, \& Gendreau}]{Hu_2023}
Hu, C.-P., Kuiper, L., Harding, A.~K., {et~al.} 2023, \bibinfo{title}{A NICER
  View on the 2020 Magnetar-like Outburst of PSR J1846â0258,} The
  Astrophysical Journal, 952, 120, \dodoi{10.3847/1538-4357/acd850}

% type= article
\bibitem[{ {Israel} {et~al.}(2007){Israel}, {Götz, D.}, {Zane, S.},
  {Dall'Osso, S.}, {Rea, N.}, \& {Stella, L.}}]{Israel_2007}
{Israel}, {Götz, D.}, {Zane, S.}, {et~al.} 2007, \bibinfo{title}{Linking the
  X-ray timing and spectral properties of the glitching AXP 1RXS
  J170849-400910â,} A\&A, 476, L9, \dodoi{10.1051/0004-6361:20078215}

% type= article
\bibitem[{G.~L. Israel {et~al.}(1994)Israel, Mereghetti, \&
  Stella}]{Israel_1994}
Israel, G.~L., Mereghetti, S., \& Stella, L. 1994, \bibinfo{title}{The
  discovery of 8.7 second pulsations from the ultrasoft X-ray source 4U
  0142+61,} \apjl, 433, L25, \dodoi{10.1086/187539}

% type= article
\bibitem[{V.~M. {Kaspi} \& A.~M. {Beloborodov}(2017){Kaspi} \&
  {Beloborodov}}]{Kaspi_2017}
{Kaspi}, V.~M., \& {Beloborodov}, A.~M. 2017, \bibinfo{title}{{Magnetars},}
  \araa, 55, 261, \dodoi{10.1146/annurev-astro-081915-023329}

% type= article
\bibitem[{V.~M. Kaspi \& F.~P. Gavriil(2003b)Kaspi \& Gavriil}]{Kaspi_2003b}
Kaspi, V.~M., \& Gavriil, F.~P. 2003b, \bibinfo{title}{A Second Glitch from the
  âAnomalousâ X-Ray Pulsar 1RXS J170849.0â4000910,} The Astrophysical
  Journal, 596, L71, \dodoi{10.1086/379093}

% type= article
\bibitem[{V.~M. Kaspi {et~al.}(2003)Kaspi, Gavriil, Woods, Jensen, Roberts, \&
  Chakrabarty}]{Kaspi_2003}
Kaspi, V.~M., Gavriil, F.~P., Woods, P.~M., {et~al.} 2003, \bibinfo{title}{A
  Major Soft Gamma Repeater-like Outburst and Rotation Glitch in the
  No-longer-so-anomalous X-Ray Pulsar 1E 2259+586,} The Astrophysical Journal,
  588, L93, \dodoi{10.1086/375683}

% type= article
\bibitem[{V.~M. Kaspi {et~al.}(2000)Kaspi, Lackey, \& Chakrabarty}]{Kaspi_2000}
Kaspi, V.~M., Lackey, J.~R., \& Chakrabarty, D. 2000, \bibinfo{title}{A Glitch
  in an Anomalous X-Ray Pulsar,} The Astrophysical Journal, 537, L31,
  \dodoi{10.1086/312758}

% type= article
\bibitem[{M.~A. Livingstone {et~al.}(2005)Livingstone, Kaspi, \&
  Gavriil}]{Livingstone_2005}
Livingstone, M.~A., Kaspi, V.~M., \& Gavriil, F.~P. 2005,
  \bibinfo{title}{Long-Term Phase-coherent X-Ray Timing of PSRÂ B0540â69,}
  The Astrophysical Journal, 633, 1095, \dodoi{10.1086/491643}

% type= article
\bibitem[{M.~E. Lower {et~al.}(2021)Lower, Johnston, Dunn, Shannon, Bailes,
  Dai, Kerr, Manchester, Melatos, Oswald, Parthasarathy, Sobey, \&
  Weltevrede}]{Lower_2021}
Lower, M.~E., Johnston, S., Dunn, L., {et~al.} 2021, \bibinfo{title}{The impact
  of glitches on young pulsar rotational evolution,} Monthly Notices of the
  Royal Astronomical Society, 508, 3251, \dodoi{10.1093/mnras/stab2678}

% type= article
\bibitem[{W.-J. Lu {et~al.}(2024)Lu, Zhou, Wang, Shao, Li, Vink, Li, \&
  Chen}]{Lu_2024}
Lu, W.-J., Zhou, P., Wang, P., {et~al.} 2024, \bibinfo{title}{Upper Limits on
  the Radio Pulses from Magnetars and a Central Compact Object with FAST,} The
  Astrophysical Journal, 963, 151, \dodoi{10.3847/1538-4357/ad27cf}

% type= article
\bibitem[{M. Lyutikov(2013)Lyutikov}]{Lyutikov_2013}
Lyutikov, M. 2013, \bibinfo{title}{Magnetospheric "anti-glitches" in
  magnetars,} arXiv, \dodoi{10.48550/arXiv.1306.2264}

% type= article
\bibitem[{E.~P. {Mazets} \& S.~V. {Golenetskii}(1979b){Mazets} \&
  {Golenetskii}}]{Mazets_b}
{Mazets}, E.~P., \& {Golenetskii}, S.~V. 1979b, \bibinfo{title}{{Observations
  of a flaring X-ray pulsar in Dorado},} Nature, 282, 287,
  \dodoi{10.1038/282587a0}

% type= article
\bibitem[{E.~P. {Mazets} {et~al.}(1979a){Mazets}, {Golenetskij}, \&
  {Guryan}}]{Mazets_a}
{Mazets}, E.~P., {Golenetskij}, S.~V., \& {Guryan}, Y.~A. 1979a,
  \bibinfo{title}{{Soft gamma-ray bursts from the source B1900+14},} Soviet
  Astronomy Letters, 5, 343

% type= article
\bibitem[{S.~A. {Olausen} \& V.~M. {Kaspi}(2014){Olausen} \&
  {Kaspi}}]{Olausen_2014}
{Olausen}, S.~A., \& {Kaspi}, V.~M. 2014, \bibinfo{title}{{The McGill Magnetar
  Catalog},} \apjs, 212, 6, \dodoi{10.1088/0067-0049/212/1/6}

% type= article
\bibitem[{H.~L. Peng {et~al.}(2024)Peng, Ge, \& Weng}]{Peng_2024}
Peng, H.~L., Ge, M.~Y., \& Weng, S.~S. 2024, \bibinfo{title}{Polarized X-Rays
  Detected from the Anomalous X-Ray Pulsar 1E 2259+586,} The Astrophysical
  Journal, 961, 106, \dodoi{10.3847/1538-4357/ad1512}

% type= article
\bibitem[{D. {Pines} \& M.~A. {Alpar}(1985){Pines} \& {Alpar}}]{Pines_1985}
{Pines}, D., \& {Alpar}, M.~A. 1985, \bibinfo{title}{Superfluidity in neutron
  stars,} Nature, 316, 27, \dodoi{10.1038/316027a0}

% type= article
\bibitem[{P.~S. Ray {et~al.}(2019)Ray, Guillot, Ho, Kerr, Enoto, Gendreau,
  Arzoumanian, Altamirano, Bogdanov, Campion, Chakrabarty, Deneva, Jaisawal,
  Kozon, Malacaria, Strohmayer, \& Wolff}]{Ray_2019}
Ray, P.~S., Guillot, S., Ho, W. C.~G., {et~al.} 2019,
  \bibinfo{title}{Anti-glitches in the Ultraluminous Accreting Pulsar NGC 300
  ULX-1 Observed with NICER,} The Astrophysical Journal, 879, 130,
  \dodoi{10.3847/1538-4357/ab24d8}

% type= inproceedings
\bibitem[{N. Rea \& P. Esposito(2011)Rea \& Esposito}]{Rea_2011}
Rea, N., \& Esposito, P. 2011, \bibinfo{title}{Magnetar outbursts: an
  observational review,} in High-Energy Emission from Pulsars and their
  Systems, ed. D.~F. Torres \& N.~Rea (Berlin, Heidelberg: Springer Berlin
  Heidelberg), 247--273

% type= article
\bibitem[{M. Rigoselli {et~al.}(2025)Rigoselli, Taverna, Mereghetti, Turolla,
  Israel, Zane, Marra, Muleri, Borghese, Coti~Zelati, De~Grandis, Imbrogno,
  Kelly, Esposito, \& Rea}]{Rigoselli_2025}
Rigoselli, M., Taverna, R., Mereghetti, S., {et~al.} 2025, \bibinfo{title}{IXPE
  Detection of Highly Polarized X-Rays from the Magnetar 1E 1841-045,} The
  Astrophysical Journal Letters, 985, L34, \dodoi{10.3847/2041-8213/adbffb}

% type= article
\bibitem[{R. Sathyaprakash {et~al.}(2024)Sathyaprakash, Rea, Coti~Zelati,
  Borghese, Pilia, Trudu, Burgay, Turolla, Zane, Esposito, Mereghetti, Campana,
  GÃ¶tz, Ibrahim, Israel, Possenti, \& Tiengo}]{Sathyaprakash_2024}
Sathyaprakash, R., Rea, N., Coti~Zelati, F., {et~al.} 2024,
  \bibinfo{title}{Long-term Study of the 2020 Magnetar-like Outburst of the
  Young Pulsar PSR J1846-0258 in Kes 75,} The Astrophysical Journal, 976, 56,
  \dodoi{10.3847/1538-4357/ad8226}

% type= article
\bibitem[{P. Scholz {et~al.}(2014)Scholz, Archibald, Kaspi, Ng, Beardmore,
  Gehrels, \& Kennea}]{Scholz_2014}
Scholz, P., Archibald, R.~F., Kaspi, V.~M., {et~al.} 2014, \bibinfo{title}{ON
  THE X-RAY VARIABILITY OF MAGNETAR 1RXSÂ J170849.0â400910,} The
  Astrophysical Journal, 783, 99, \dodoi{10.1088/0004-637X/783/2/99}

% type= article
\bibitem[{R. Stewart {et~al.}(2025)Stewart, Younes, Harding, Wadiasingh,
  Baring, Negro, Strohmayer, Ho, Ng, Arzoumanian, Thi, Di~Lalla, Enoto,
  Gendreau, Hu, van Kooten, Kouveliotou, \& McEwen}]{Stewart_2025}
Stewart, R., Younes, G.~A., Harding, A.~K., {et~al.} 2025,
  \bibinfo{title}{X-Ray Polarization of the Magnetar 1E 1841â045,} The
  Astrophysical Journal Letters, 985, L35, \dodoi{10.3847/2041-8213/adbffa}

% type= article
\bibitem[{M. Sugizaki {et~al.}(1997)Sugizaki, Nagase, Torii, Kinugasa, Asanuma,
  Matsuzaki, Koyama, \& Yamauchi}]{Sugizaki_1997}
Sugizaki, M., Nagase, F., Torii, K., {et~al.} 1997, \bibinfo{title}{Discovery
  of an 11-s X-Ray Pulsar in the Galactic-Plane Section of the Scorpius
  Constellation,} Publications of the Astronomical Society of Japan, 49, L25,
  \dodoi{10.1093/pasj/49.5.L25}

% type= article
\bibitem[{C. Thompson \& R.~C. Duncan(1995)Thompson \& Duncan}]{Thompson_1995}
Thompson, C., \& Duncan, R.~C. 1995, \bibinfo{title}{The soft gamma repeaters
  as very strongly magnetized neutron stars - I. Radiative mechanism for
  outbursts,} Monthly Notices of the Royal Astronomical Society, 275, 255,
  \dodoi{10.1093/mnras/275.2.255}

% type= article
\bibitem[{C. Thompson \& R.~C. Duncan(1996)Thompson \& Duncan}]{Thompson_1996}
Thompson, C., \& Duncan, R.~C. 1996, \bibinfo{title}{The Soft Gamma Repeaters
  as Very Strongly Magnetized Neutron Stars. II. Quiescent Neutrino, X-Ray, and
  AlfvÃ©n Wave Emission,} The Astrophysical Journal, 473, 322,
  \dodoi{10.1086/178147}

% type= article
\bibitem[{C. {Thompson} {et~al.}(2000){Thompson}, {Duncan}, {Woods},
  {Kouveliotou}, {Finger}, \& {van Paradijs}}]{Thompson_2000}
{Thompson}, C., {Duncan}, R.~C., {Woods}, P.~M., {et~al.} 2000,
  \bibinfo{title}{{Physical Mechanisms for the Variable Spin-down and Light
  Curve of SGR 1900+14},} \apj, 543, 340, \dodoi{10.1086/317072}

% type= article
\bibitem[{H. Tong(2014)Tong}]{Tong_2014}
Tong, H. 2014, \bibinfo{title}{THE ANTI-GLITCH OF MAGNETAR 1E 2259+586 IN THE
  WIND BRAKING SCENARIO,} The Astrophysical Journal, 784, 86,
  \dodoi{10.1088/0004-637X/784/2/86}

% type= misc
\bibitem[{H. Tong(2023)Tong}]{tong_2023}
Tong, H. 2023, Magnetospheric physics of magnetars, \doarXiv{2309.05181}

% type= article
\bibitem[{Y.~L. Tuo {et~al.}(2024)Tuo, Serim, Antonelli, Ducci, Vahdat, Ge,
  Santangelo, \& Xie}]{Tuo_2024}
Tuo, Y.~L., Serim, M.~M., Antonelli, M., {et~al.} 2024,
  \bibinfo{title}{Discovery of the First Antiglitch Event in the
  Rotation-powered Pulsar PSR B0540-69,} The Astrophysical Journal Letters,
  967, L13, \dodoi{10.3847/2041-8213/ad4488}

% type= article
\bibitem[{R. Turolla {et~al.}(2015)Turolla, Zane, \& Watts}]{Turolla_2015}
Turolla, R., Zane, S., \& Watts, A.~L. 2015, \bibinfo{title}{Magnetars: the
  physics behind observations. A review,} Reports on Progress in Physics, 78,
  116901, \dodoi{10.1088/0034-4885/78/11/116901}

% type= article
\bibitem[{G. Vasisht \& E.~V. Gotthelf(1997)Vasisht \& Gotthelf}]{Vasisht_1997}
Vasisht, G., \& Gotthelf, E.~V. 1997, \bibinfo{title}{The Discovery of an
  Anomalous X-Ray Pulsar in the Supernova Remnant Kes 73,} The Astrophysical
  Journal, 486, L129, \dodoi{10.1086/310843}

% type= article
\bibitem[{T. Wong {et~al.}(2001)Wong, Backer, \& Lyne}]{Wong_2001}
Wong, T., Backer, D.~C., \& Lyne, A.~G. 2001, \bibinfo{title}{Observations of a
  Series of Six Recent Glitches in the Crab Pulsar,} The Astrophysical Journal,
  548, 447, \dodoi{10.1086/318657}

% type= article
\bibitem[{P.~M. Woods {et~al.}(2004)Woods, Kaspi, Thompson, Gavriil, Marshall,
  Chakrabarty, Flanagan, Heyl, \& Hernquist}]{Woods_2004}
Woods, P.~M., Kaspi, V.~M., Thompson, C., {et~al.} 2004,
  \bibinfo{title}{Changes in the X-Ray Emission from the Magnetar Candidate 1E
  2259+586 during Its 2002 Outburst,} The Astrophysical Journal, 605, 378,
  \dodoi{10.1086/382233}

% type= article
\bibitem[{G. Younes {et~al.}(2023)Younes, Baring, Harding, \&
  Enoto}]{Younes_2023}
Younes, G., Baring, M.~G., Harding, A.~K., \& Enoto, T. 2023,
  \bibinfo{title}{Magnetar spin-down glitch clearing the way for FRB-like
  bursts and a pulsed radio episode,} Nature Astronomy, 7, 339,
  \dodoi{10.1038/s41550-022-01865-y}

% type= article
\bibitem[{G. Younes {et~al.}(2025)Younes, Lander, \& Baring}]{Younes_2025}
Younes, G., Lander, S.~K., \& Baring, M.~G. 2025, \bibinfo{title}{Timing and
  Spectral Evolution of the Magnetar 1E 1841-045 in Outburst,} arXiv,
  \dodoi{10.48550/arXiv.2502.20079}

% type= article
\bibitem[{G. Younes {et~al.}(2020)Younes, Ray, Baring, Kouveliotou, Fletcher,
  Wadiasingh, Harding, \& Goldstein}]{Younes_2020}
Younes, G., Ray, P.~S., Baring, M.~G., {et~al.} 2020, \bibinfo{title}{A
  Radiatively Quiet Glitch and Anti-glitch in the Magnetar 1EÂ 2259+586,} The
  Astrophysical Journal Letters, 896, L42, \dodoi{10.3847/2041-8213/ab9a48}

% type= article
\bibitem[{S.-N. {Zhang} {et~al.}(2025){Zhang}, {Santangelo}, {Xu}, {Feng},
  {Lu}, {Chen}, {Ge}, {Nandra}, {Wu}, {Feroci}, {Hernanz}, {Liu}, {He}, {Wang},
  {Jiang}, {Cui}, {Yang}, {Wang}, {Li}, {Liu}, {Meng}, {Wen}, {Zhang}, {Ma},
  {Li}, {Li}, {Qi}, {Sun}, {Luo}, {Liu}, {Liu}, {Zhang}, {Luo}, {Zhu}, {Zhao},
  {Sun}, {Yang}, {Wu}, {Jiang}, {Shi}, {Liu}, {Xu}, {Yang}, {Zhang}, {Han},
  {Gao}, {Huo}, {Zhang}, {Wang}, {Zhao}, {Cui}, {Wang}, {Wang}, {Li}, {Bao},
  {Liu}, {Wang}, {Wang}, {Wang}, {Wang}, {Wang}, {Ding}, {Sheng}, {Qiang},
  {Yan}, {Liu}, {Wu}, {Liu}, {Chen}, {Zhang}, {Liu}, {Altmann}, {Bechteler},
  {Burwitz}, {Fiorini}, {Friedrich}, {Meidinger}, {Strecker}, {Baldini},
  {Bellazzini}, {Bonino}, {Frass{\`a}}, {Latronico}, {Maldera}, {Manfreda},
  {Minuti}, {Pesce-Rollins}, {Sgr{\`o}}, {Tugliani}, {Pareschi}, {Basso},
  {Sironi}, {Spiga}, {Tagliaferri}, {Tykhonov}, {Paltani}, {Bozzo}, {Tenzer},
  {Bayer}, {Tuo}, {Liu}, {Zhang}, {Cai}, {Liu}, {Chen}, {Wang}, {He}, {Chen},
  {Qiu}, {Zhang}, {Feng}, {Zhu}, {Zhou}, {Zheng}, {Song}, {Shi}, {Wang}, {Jia},
  {Jiang}, {Li}, {Zhao}, {Guan}, {Zhang}, {Li}, {Huang}, {Liao}, {You},
  {Zhang}, {Wang}, {Wang}, {Ou}, {Hu}, {Shi}, {Cui}, {Jiang}, {Cheng}, {Li},
  {Xu}, {Zane}, {Bambi}, {Bu}, {Dall'Osso}, {De Rosa}, {Gou}, {Guillot}, {Ji},
  {Li}, {Mao}, {Patruno}, {Stratta}, {Taverna}, {Tsygankov}, {Uttley}, {Watts},
  {Wu}, {Xu}, {Yi}, {Zhang}, {Zhang}, {Zhao}, \& {Zhou}}]{Zhang_2025}
{Zhang}, S.-N., {Santangelo}, A., {Xu}, Y., {et~al.} 2025, \bibinfo{title}{{The
  enhanced X-ray Timing and Polarimetry mission -- eXTP for launch in 2030},}
  arXiv e-prints, arXiv:2506.08101, \dodoi{10.48550/arXiv.2506.08101}

% type= article
\bibitem[{S.~Q. Zhou {et~al.}(2022a)Zhou, Gügercinoğlu, Yuan, Ge, \&
  Yu}]{zhou_2022a}
Zhou, S.~Q., Gügercinoğlu, E., Yuan, J., Ge, M., \& Yu, C. 2022a,
  \bibinfo{title}{Pulsar Glitches: A Review,} Universe, 8,
  \dodoi{10.3390/universe8120641}

% type= article
\bibitem[{S.~Q. Zhou {et~al.}(2024)Zhou, Ye, Ge, Gügercinoğlu, Zheng, Yu,
  Yuan, \& Zhang}]{Zhou_2024}
Zhou, S.~Q., Ye, W.~T., Ge, M.~Y., {et~al.} 2024, \bibinfo{title}{A Series of
  (Net) Spin-down Glitches in PSR J1522â5735: Insights from the Vortex Creep
  and Vortex Bending Models,} The Astrophysical Journal, 977, 243,
  \dodoi{10.3847/1538-4357/ad938d}

% type= article
\bibitem[{S.~Q. Zhou {et~al.}(2022b)Zhou, Gügercinoğlu, Yuan, Ge, Yu, Zhang,
  Zhang, Feng, \& Ye}]{zhou_2022b}
Zhou, S.~Q., Gügercinoğlu, E., Yuan, J.~P., {et~al.} 2022b,
  \bibinfo{title}{New pulse profile variability associated with a glitch of PSR
  J0738-4042,} Monthly Notices of the Royal Astronomical Society, 519, 74,
  \dodoi{10.1093/mnras/stac3355}

% type= article
\bibitem[{B. İçdem {et~al.}(2012)Icdem, Baykal, \& İnam}]{Icdem_2012}
İçdem, B., Baykal, A., \& İnam,  S. Ç. 2012, \bibinfo{title}{RXTE timing
  analysis of the anomalous X-ray pulsar 1E 2259+586,} Monthly Notices of the
  Royal Astronomical Society, 419, 3109,
  \dodoi{10.1111/j.1365-2966.2011.19953.x}

% type= article
\bibitem[{S. Şaşmaz Muş {et~al.}(2014)Şaşmaz Muş, Aydın, \&
  Göğüş}]{Sinem_2014}
Şaşmaz Muş, S., Aydın, B., \& Göğüş, E. 2014, \bibinfo{title}{A glitch
  and an anti-glitch in the anomalous X-ray pulsar 1E 1841â045,} Monthly
  Notices of the Royal Astronomical Society, 440, 2916,
  \dodoi{10.1093/mnras/stu436}

\end{thebibliography}
%\bibliographystyle{aasjournalv7}

\end{document}